\documentclass[12pt]{article}
\usepackage{graphicx}
\usepackage{amssymb}

\begin{document}

\newcommand{\beq}{\begin{equation}}
\newcommand{\eeq}{\end{equation}}
\newcommand{\barr}{\begin{eqnarray}}
\newcommand{\earr}{\end{eqnarray}}

\newcommand{\andy}[1]{ }

\def\cH{{\mathcal{H}}}
\def\cV{{\mathcal{V}}}
\def\cU{{\mathcal{U}}}
\def\cP{{\mathcal{P}}}
\def\cN{{\mathcal{N}}}
\def\cS{{\mathcal{S}}}
\def\As{{\mathcal{A}}}
\def\bra#1{\langle #1 |}
\def\ket#1{| #1 \rangle}
\def\bbra#1{( #1 |}
\def\kket#1{| #1 )}
\def\Ord{\mathrm{O}}
\renewcommand{\Re}{\mathrm{\,Re\,}}
\renewcommand{\Im}{\mathrm{\,Im\,}}

\title{Quantum Zeno subspaces and dynamical superselection rules}

\author{Paolo Facchi and Saverio Pascazio \\
\quad \\ Dipartimento di Fisica, Universit\`a di Bari
\\
and Istituto Nazionale di Fisica Nucleare, Sezione di Bari\\
I-70126 Bari, Italy }

\maketitle

\abstract{ The quantum Zeno evolution of a quantum system
takes place in a proper subspace of the total Hilbert space. The
physical and mathematical features of the ``Zeno subspaces" depend
on the measuring apparatus: when this is included in the quantum
description, the Zeno effect becomes a mere consequence of the
dynamics and, remarkably, can be cast in terms of an adiabatic
theorem, with a dynamical superselection rule. We look at several
examples and focus on quantum computation and decoherence-free
subspaces.}


\section{Introduction }
 \label{sec-introd}
 \andy{intro}

The quantum Zeno effect has a curious history. It was first
understood by von Neumann, in 1932 \cite{von}: while analyzing the
thermodynamic features of quantum ensembles, at page 195 of his
book on the Mathematical Foundations of Quantum Mechanics (page
366 of the English translation), von Neumann proved that any given
state $\phi$ of a quantum mechanical system can be ``steered" into
any other state $\psi$ of the same Hilbert space, by performing a
series of very frequent measurements. If $\phi$ and $\psi$
coincide (modulo a phase factor), the evolution is ``frozen" and,
in modern language, a quantum Zeno effect takes place.

This remarkable observation did not trigger much interest, neither
in the mathematical, nor in the physical literature. It took 35
years before Beskow and Nilsson \cite{Beskow} applied the same
ideas to a rather concrete physical problem (a particle in a
bubble chamber) and wondered whether it is possible to influence
the decay of an unstable system by performing frequent
``observations" on it (a bubble chamber can be thought of as an
apparatus that ``continuously" checks whether the particle has
decayed). This interesting idea was subsequently physically
analyzed by several authors
\cite{Khalfin68,Misra,Peres80,KrausSud}.
The classical allusion to the sophist philosopher Zeno of Elea is
due to Misra and Sudarshan \cite{Misra}, who were also the first
to provide a consistent and rigorous mathematical framework.
During those years it was also realized that the formulation of
the ``Zeno effect" (or ``paradox" as people tended to regard it)
hinged upon difficult mathematical issues
\cite{Friedman72,Gustafson,GustafsonSolvay}, most of which are yet unsolved.

The interest in the quantum Zeno effect (QZE) was revived in 1988,
when Cook \cite{Cook} proposed to test it on oscillating (mainly,
two-level) systems, rather than on \textit{bona fide} unstable
ones. This was an interesting and concrete idea, that led to
experimental test a few years later \cite{Itano}. The discussion
that followed \cite{Itanodisc,PN94} provided alternative insight
and new ideas \cite{varieidee}, eventually leading to new
experimental tests. The QZE was successfully checked in
experiments involving photon polarization \cite{kwiat}, chiral
molecules \cite{Chapovsky} and ions \cite{Balzer} and new
experiments are in preparation with neutron spin \cite{VESTA}. One
should emphasize that the first experiments were not free from
interpretational criticisms. Some of these criticisms could be
successfully countered (e.g., the serious problem related to the
so-called ``repopulation" of the initial state
\cite{Nakazato96b,zenoreview} was avoided in \cite{Balzer}), but
some authors insisted in arguing that the QZE had not been
successfully demonstrated on \emph{bona fide} unstable systems, as
in the seminal proposals.

Fortunately (or unfortunately, depending on the perspective) the
recent experiments by Raizen and collaborators are conclusive, in
our opinion:  the presence of a short-time quadratic region for an
unstable quantum mechanical system (particle tunnelling out of a
confining potential) was experimentally confirmed in 1997
\cite{Wilkinson} and then, a few years later, the existence of the
Zeno effect (hindered evolution by frequent measurements) was
demonstrated \cite{raizenlatest}. This last experiment is of great
conceptual interest, for it also proved the occurrence of the
so-called inverse (or anti) Zeno effect (IZE)
\cite{antiZeno,heraclitus,zenowaseda}), first suggested in
1983~(!), according to which the evolution can be
\emph{accelerated} if the measurements are frequent, but not
\emph{too} frequent.

The QZE is a direct consequence of general features of the
Schr\"odinger equation that yield quadratic behavior of the
survival probability at short times \cite{strev,zenoreview}.
According to the standard formulation, the hindrance of the
evolution is due to very frequent measurements, aimed at
ascertaining whether the quantum system is still in its initial
state. We call this a ``pulsed measurement" formulation
\cite{zenoreview}, according to von Neumann's projection postulate
\cite{von}. However, from a physical point of view, a
``measurement" is nothing but an interaction with an external
system (another quantum object, or a field, or simply a different
degree of freedom of the very system investigated), playing the
role of apparatus. If the apparatus is included in the quantum
description, the QZE can be reformulated in terms of a
``continuous" measurement \cite{zenoreview,Napoli,zenowaseda},
without making use of projection operators and non-unitary
dynamics, obtaining the same physical effects. It is important to
stress that the idea of a ``continuous" formulation of the QZE is
not new \cite{Peres80,KrausSud}, but a \emph{quantitative}
comparison with the ``pulsed" situation is rather recent
\cite{Schulman98}.

Nowadays, it seems therefore more appropriate to frame the Zeno
effects in a dynamical scenario \cite{PN94} by making use of a
continuous-measurement formulation
\cite{zenoreview,Napoli,Schulman98,varicont,Panov}. Also, it is
important to focus on additional issues, in view of possible
applications. For instance, it is interesting to notice that a
quantum Zeno evolution does not necessarily freeze the dynamics.
On the contrary, for frequent projections onto a multidimensional
subspace, the system can evolve away from its initial state,
although it remains in the subspace defined by the ``measurement"
\cite{compactregularize}. By blending together these three
ingredients (dynamical framework, continuous measurement and Zeno
dynamics within a subspace) the quantum Zeno evolution can be cast
in terms of an adiabatic theorem \cite{theorem}: under the action
of a continuous measurement process (and in a strong coupling
limit to be defined in the following) the system is forced to
evolve in a set of orthogonal subspaces of the total Hilbert space
and an \emph{effective superselection rule} arises. The
dynamically disjoint \textit{quantum Zeno subspaces} are the
eigenspaces (belonging to different eigenvalues) of the
Hamiltonian that describes the interaction between the system and
the apparatus: in words, they are those subspaces that the
measurement device is able to distinguish.

This paves the way to possible interesting applications of the
QZE: indeed, if the coupling between the ``observed" system and
the ``measuring" apparatus can be tailored in order to slow (or
accelerate) the evolution, a door is open to control unwanted
effects, such as decoherence and dissipation. It is therefore
important to understand in great detail when an external quantum
system can be considered a good ``apparatus," able to yield QZE
and IZE, and why.

We have organized our discussion as follows. We first review in
Sec.\ \ref{sec-dpw} some notions related to the (familiar)
``pulsed" formulation of the Zeno effect and summarize the
celebrated Misra and Sudarshan theorem in Sec.\ \ref{sec-msth}.
This theorem is then extended in Sec.\ \ref{sec-partial}, in order
to accommodate multiple projectors, and the notion of continuous
measurement is introduced in Sec.\ \ref{sec-QZEcont}, by looking
at several examples. We propose in Sec.\ \ref{sec-novdef} a
broader definition of QZE (and IZE) \cite{zenoreview} and prove in
Sec.\ \ref{sec-dynamicalQZE} an adiabatic theorem, defining the
Zeno subspaces \cite{theorem,thesis}. Finally, in Secs.\
\ref{sec-applications}-\ref{sec-vacdec}, we elaborate on some
interesting examples, focusing in particular on quantum
computation and applications. We conclude in Sec.\ \ref{sec-concl}
with a few comments.

\section{Notation and preliminary notions:\\ pulsed measurements}
\label{sec-dpw}
\andy{sec-dpw}

Let $H$ be the total Hamiltonian of a quantum system and $\ket{a}$
its initial state at $t=0$. The survival probability in state
$\ket{a}$ is
\andy{uno}
\beq
p(t) = |\As (t)|^2 =|\langle a|e^{-iHt}|a\rangle |^2
\label{eq:uno}
\eeq
and a short-time expansion yields a quadratic behavior
\andy{quadratic}
\beq
p(t) \sim 1 - t^2/\tau_{\mathrm{Z}}^2, \qquad
\tau_{\mathrm{Z}}^{-2} \equiv \langle a|H^2|a\rangle - \langle
a|H|a\rangle^2 ,
\label{eq:quadratic}
\eeq
where $\tau_{\mathrm{Z}}$ is the Zeno time \cite{hydrogen}.
Observe that if the Hamiltonian is divided into a free and an
(off-diagonal) interaction parts
\andy{Hdiv}
\beq
H=H_0 + H_{\mathrm{int}}, \qquad \mbox{with} \quad
H_0\ket{a}=\omega_a\ket{a}, \quad
\bra{a}H_{\mathrm{int}}\ket{a}=0, \label{eq:Hdiv}
\eeq
the Zeno time reads
\andy{tzoff}
\beq
\tau_{\mathrm{Z}}^{-2} = \bra{a}H_{\mathrm{int}}^2\ket{a}
\label{eq:tzoff}
\eeq
and depends only on the interaction Hamiltonian.

\begin{figure}[t]
\begin{center}
\includegraphics[width=6cm]{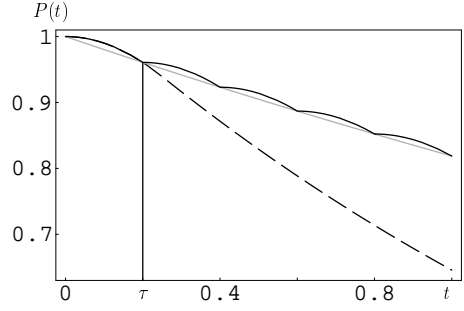}
\end{center}
\caption{Evolution with frequent ``pulsed" measurements: quantum
Zeno effect. The dashed (full) line is the survival probability
without (with) measurements. The gray line is the interpolating
exponential (\ref{eq:survN0}).} \label{fig:zenoevol}
\end{figure}

Perform $N$ (instantaneous) measurements at time intervals
$\tau=t/N$, in order to check whether the system is still in state
$\ket{a}$. The survival probability after the measurements reads
\andy{survN}
\beq
p^{(N)}(t)=p(\tau)^N = p\left(t/N\right)^N
\sim\exp\left(-t^2/\tau_{\mathrm{Z}}^2 N\right) \stackrel{N
\rightarrow\infty}{\longrightarrow} 1 .
\eeq
If $N=\infty$ the evolution is completely hindered. For very large
(but finite) $N$ the evolution is slowed down: indeed, the
survival probability after $N$ pulsed measurements ($t=N\tau$) is
interpolated by an exponential law \cite{heraclitus}
\andy{survN0}
\beq
p^{(N)}(t)=p(\tau)^N=\exp(N\log p(\tau))=
\exp(-\gamma_{\mathrm{eff}}(\tau) t) ,
\label{eq:survN0}
\eeq
with an \emph{effective decay rate}
\andy{eq:gammaeffdef}
\beq
\gamma_{\mathrm{eff}}(\tau) \equiv -\frac{1}{\tau}\log p(\tau) =
-\frac{2}{\tau}\log |\As (\tau)| =-\frac{2}{\tau}\Re [\log
\As(\tau)] \ge0 \; . \label{eq:gammaeffdef}
\eeq
For $\tau\to 0 $ (i.e.\ $N \to \infty$) one gets  $p(\tau) \sim
\exp (-\tau^2/\tau_{\mathrm{Z}}^2)$, whence
\beq
\gamma_{\mathrm{eff}}(\tau) \sim \tau/\tau_{\mathrm{Z}}^2. \qquad
(\tau\to 0)
\label{eq:lingammaeff}
\eeq
Increasingly frequent measurements tend to hinder the evolution.
The \emph{physical} meaning of the mathematical expression
``$\tau\to 0$" is a subtle issue
\cite{hydrogen,heraclitus,zenoreview,Antoniou}, involving quantum
field theoretical considerations \cite{QFT,Panov,zenowaseda} that
will not be considered here. The Zeno evolution for ``pulsed"
measurements is pictorially represented in Figure
\ref{fig:zenoevol}. The notion of ``continuous" measurement will
be discussed later (Sec.\ \ref{sec-QZEcont}).

\section{Misra and Sudarshan's theorem}
\label{sec-msth}
\andy{msth}

We briefly sketch Misra and Sudarshan's theorem and introduce more
notation. Let Q be a quantum system, whose states belong to the
Hilbert space ${\cal H}$ and whose evolution is described by the
unitary operator $U(t)=\exp(-iHt)$, where $H$ is a
time-independent lower-bounded Hamiltonian. Let $P$ be a
projection operator and $\textrm{Ran}P=\cH_P$ its range. We assume
that the initial density matrix $\rho_0$ of system Q belongs to
${\cal H}_P$:
\andy{inprep}
\beq
\rho_0 = P \rho_0 P , \qquad \mathrm{Tr} [ \rho_0 P ] = 1 .
\label{eq:inprep}
\eeq
Under the action of the Hamiltonian $H$ (i.e., if no measurements
are performed in order to get information about the quantum
state), the state at time $t$ reads
\andy{noproie}
\beq
\rho (t) = U(t) \rho_0 U^\dagger (t)
  \label{eq:noproie}
\eeq
and the \textit{survival probability}, namely the probability that
the system is still in ${\cal H}_P$ at time $t$, is
\andy{stillun}
\beq
p(t) = \mathrm{Tr} \left[ U(t) \rho_0 U^\dagger(t) P \right] .
\label{eq:stillun}
\eeq
No distinction is made between one- and multi-dimensional
projections.

The above evolution is ``undisturbed," in the sense that the
quantum systems evolves only under the action of its Hamiltonian
for a time $t$, without undergoing any measurement process.
Assume, on the other hand, that we do perform a \textit{selective
measurement} at time $\tau$, in order to check whether Q has
survived inside ${\cal H}_P$. By this, we mean that we select the
survived component and stop the other ones. (Think for instance of
spectrally decomposing a spin in a Stern-Gerlach setup and
absorbing away the unwanted components.)

The state of Q changes (up to a normalization constant) into
\andy{proie}
\beq
\rho_0 \rightarrow \rho(\tau) = P U(\tau) \rho_0 U^\dagger(\tau) P
\label{eq:proie}
\eeq
and the survival probability in ${\cal H}_P$ is
\andy{probini}
\barr
p(\tau) = \mathrm{Tr} \left[ U(\tau) \rho_0 U^\dagger(\tau) P
\right] = \mathrm{Tr} \left[V(\tau) \rho_0 V^\dagger(\tau)
\right],
  \qquad V(\tau) \equiv P U(\tau)P.
\label{eq:probini}
\earr
The QZE is the following. We prepare Q in the initial state
$\rho_0$ at time 0 and perform a series of (selective)
$P$-observations at time intervals $\tau=t/N$. The state of Q at
time $t$ reads (up to a normalization constant)
\andy{Nproie}
\beq
\rho^{(N)}(t) = V_N(t) \rho_0 V_N^\dagger(t) , \qquad
    V_N(t) \equiv [ P U(t/N) P ]^N
\label{eq:Nproie}
\eeq
and the survival probability in ${\cal H}_P$ is given by
\andy{probNob}
\beq
p^{(N)}(t) = \mathrm{Tr} \left[ V_N(t) \rho_0 V_N^\dagger(t)
\right]. \label{eq:probNob}
\eeq
In order to consider the $N \rightarrow \infty$ limit, one needs
some mathematical requirements: assume that the limit
\andy{slim}
\beq
\cV (t) \equiv \lim_{N \rightarrow \infty} V_N(t)
  \label{eq:slim}
\eeq
exists (in the strong sense) for $t>0$. The final state of Q is
then
\andy{infproie}
\beq
\rho (t) = \lim_{N\to\infty} \rho^{(N)}(t)=\cV(t) \rho_0
\cV^\dagger (t)
  \label{eq:infproie}
\eeq
and the probability to find the system in $\cH_P$ is
\andy{probinfob}
\beq
\cP (t) \equiv \lim_{N \rightarrow \infty} p^{(N)}(t)
   = \mathrm{Tr} \left[ \cV(t) \rho_0 \cV^\dagger(t) \right].
\label{eq:probinfob}
\eeq
By assuming the strong continuity of $\cV(t)$ at $t=0$
\andy{phgr}
\beq
\lim_{t \rightarrow 0^+} \cV(t) = P, \label{eq:phgr}
\eeq
Misra and Sudarshan proved that under general conditions the
operators
\andy{semigr}
\beq
\cV(t) \quad \textrm{exist for all real $t$ and form a semigroup.}
\label{eq:semigr}
\eeq
Moreover, by time-reversal invariance
\andy{VVdag}
\beq
\cV^\dagger (t) = \cV(-t), \label{eq:VVdag}
\eeq
one gets $\cV^\dagger (t) \cV(t) =P$. This implies, by
(\ref{eq:inprep}), that
\andy{probinfu1}
\beq
\cP(t)=\mathrm{Tr}\left[\rho_0 \cV^\dagger(t)\cV(t)\right] =
\mathrm{Tr} \left[ \rho_0 P \right] = 1 . \label{eq:probinfu1}
\eeq
If the particle is very frequently observed, in order to check
whether it has survived inside $\cH_P$, it will never make a
transition to $\cH_P^\perp$ (QZE). In general, if $N$ is
sufficiently large in (\ref{eq:Nproie})-(\ref{eq:probNob}), all
transitions outside ${\cal H}_P$ are inhibited.

We emphasize that close scrutiny of the features of the survival
probability has clarified that if $N$ is not \emph{too} large the
system can display an inverse Zeno effect
\cite{antiZeno,heraclitus,zenowaseda}, by which decay is
accelerated. Both effects have recently been seen in the same
experimental setup \cite{raizenlatest}. We will not elaborate on
this here.

Notice also that the dynamics (\ref{eq:Nproie})-(\ref{eq:probNob})
is not reversible. On the other hand, the dynamics in the $N \to
\infty$ limit is often time reversible \cite{compactregularize}
(although, in general, the operators $\cV(t)$ in (\ref{eq:semigr})
form a \textit{semi}group).

The theorem just summarized \textit{does not} state that the
system \textit{remains} in its initial state, after the series of
very frequent measurements. Rather, the system \textit{evolves} in
the subspace $\cH_P$, instead of evolving ``naturally" in the
total Hilbert space $\cH$. The features of this evolution will be
the object study of the following sections.

\section{Multidimensional measurements}
 \label{sec-partial}
 \andy{partial}

We now analyze the (most interesting) case of multidimensional
measurements. We will apply the von Neumann-L\"{u}ders
\cite{von,Luders} formulation in terms of projection operators, by
adopting some definitions given by Schwinger \cite{Schwinger59}.

\subsection{Incomplete measurements}
 \label{sec-incomplete}
 \andy{incomplete}

We will say that a measurement is ``incomplete" if some outcomes
are lumped together. This happens, for example, if the
experimental equipment has insufficient resolution (and in this
sense the information on the measured observable is
``incomplete"). See, for example, \cite{Peres98}. The projection
operator $P$, which selects a particular lump, is therefore
multidimensional. Let us first consider a \emph{finite}
dimensional $\cH_P=\mathrm{Ran}P$,
\beq
\mathrm{dim} \cH_P = \mathrm{Tr} P=s < \infty .
\eeq
The resulting time evolution operator is a finite dimensional
matrix and has the explicit form
\beq
\cV (t)= \lim_{N\to\infty} V_N(t) = \lim_{N\to\infty} [ P U(t/N) P
]^N =P \exp(-i P H P t). \label{eq:cVfin1}
\eeq
It is easy to show that if $\cH_P\subset D(H)$, the domain of the
Hamiltonian $H$, then $\cV(t)$ in (\ref{eq:cVfin1}) is unitary
within $\cH_P$ and is generated by the self-adjoint Hamiltonian
$PHP$ (an example is given in \cite{MNPRY}). Reversibility is
recovered in the $N \to \infty$ limit.

For infinite dimensional projections, $s=\infty$, one can always
formally write the limiting evolution in the form
(\ref{eq:cVfin1}), but has to define the meaning of $P H P$. In
such a case the time evolution operator $\cV(t)$ may be not
unitary and one has to study the self-adjointness of the limiting
Hamiltonian $PHP$
\cite{compactregularize,Friedman72,Gustafson,GustafsonSolvay}.

In general, for incomplete measurements, system Q does \emph{not}
remain in its initial state. Rather, it is confined in the
subspace $\cH_P$ and evolves under the action of $\cV(t)$, instead
of evolving ``naturally" in the total Hilbert space $\cH$.

\subsection{Nonselective measurements}
 \label{sec-nonselect}
 \andy{nonselect}

We will say that a measurement is ``nonselective"
\cite{Schwinger59} if the measuring apparatus does not ``select"
the different outcomes, so that all the ``beams" (after the
spectral decomposition
\cite{Wigner63,PN94,NPN}) undergo the whole Zeno dynamics. In
other words, a nonselective measurement destroys the phase
correlations between different branch waves, provoking the
transition from a pure state to a mixture.

We now consider the case of nonselective measurements and extend
Misra and Sudarshan's theorem in order to accommodate multiple
projectors and build a bridge for our subsequent discussion. Let
\beq
\{P_n\}_n, \qquad
P_nP_m=\delta_{mn}P_n,\qquad  \sum_n P_n=1 ,
\eeq
be a (countable) collection of projection operators and
$\mathrm{Ran}P_n=\cH_{P_n}$ the relative subspaces. This induces a
partition on the total Hilbert space
\beq
\label{eq:partition}
\cH=\bigoplus_n \cH_{P_n}.
\eeq
Consider the associated nonselective measurement described by the
superoperator \cite{von,Luders}
\beq
\label{eq:superP} \hat P \rho=\sum_n P_n \rho P_n.
\eeq
The free evolution reads
\beq
\hat U_t \rho_0=U(t) \rho_0 U^\dagger(t),\qquad U(t)=\exp(-i H t)
\eeq
and the Zeno evolution after $N$ measurements in a time $t$ is
governed by the superoperator
\beq
\hat V^{(N)}_t=\hat P\left(\hat U \left(t/N \right)\hat
P\right)^{N-1} .
\eeq
This yields the evolution
\beq
\rho(t)=\hat V^{(N)}_t \rho_0 =\sum_{n_1,\dots,n_N}V_{n_1\dots
n_N}^{(N)}(t)\; \rho_0\; V_{n_1\dots n_N}^{(N)\dagger}(t) ,
\eeq
where
\barr
V_{n_1\dots n_N}^{(N)}(t)  = P_{n_N} U\left(t/N\right) P_{n_{N-1}}
\cdots P_{n_2} U\left(t/N\right) P_{n_1}, \label{eq:boo}
\earr
which should be compared to Eq.\ (\ref{eq:Nproie}). We follow
Misra and Sudarshan \cite{Misra} and assume, as in Sec.\
\ref{sec-msth}, the time-reversal invariance and the existence of
the strong limits ($t>0$)
\andy{slims}
\beq
\cV_n (t)=\lim_{N\to\infty} V_{n\dots n}^{(N)}(t) , \qquad \lim_{t
\rightarrow 0^+} \cV_n(t) = P_n , \quad \forall n \ .
\label{eq:slims}
\eeq
Then $\cV_n(t)$ exist for all real $t$ and form a semigroup
\cite{Misra}, and
\beq
\cV_n^\dagger(t)\cV_n(t)=P_n.
\eeq
Moreover, it is easy to show that
\andy{fulldiag}
\beq
\lim_{N\to\infty} V_{n\dots n'\dots}^{(N)}(t) = 0, \qquad
\mathrm{for}\quad n'\neq n .
\label{eq:fulldiag}
\eeq
Notice that, for any \emph{finite} $N$, the off-diagonal operators
(\ref{eq:boo}) are in general nonvanishing, i.e.\ $V_{n\dots
n'\dots}^{(N)}(t) \neq 0$ for $n'\neq n$. It is only in the limit
(\ref{eq:fulldiag}) that these operators become diagonal. This is
because $U\left(t/N\right)$ provokes transitions among different
subspaces $\cH_{P_n}$. By Eqs.\
(\ref{eq:slims})-(\ref{eq:fulldiag}) the final state is
\andy{rhoZ}
\barr
\rho(t)=\hat \cV_t\rho_0 =\sum_n \cV_n(t) \rho_0 \cV_n^\dagger(t),
\quad \mathrm{with} \quad \sum_n \cV_n^\dagger(t)\cV_n(t)=\sum_n
P_n=1 .\;\; \label{eq:rhoZ}
\earr
The components $\cV_n(t) \rho_0 \cV_n^\dagger(t)$ make up a block
diagonal matrix: the initial density matrix is reduced to a
mixture and any interference between different subspaces
$\cH_{P_n}$ is destroyed (complete decoherence). In conclusion,
\andy{probinfu}
\beq
p_n(t) =  \mathrm{Tr} \left[\rho(t) P_n\right]=
\mathrm{Tr}\left[\rho_0 P_n\right]=p_n(0) , \quad \forall n .
\label{eq:probinfu}
\eeq
In words, probability is conserved in each subspace and no
probability ``leakage" between any two subspaces is possible: the
total Hilbert space splits into invariant subspaces and the
different components of the wave function (or density matrix)
evolve independently within each sector. One can think of the
total Hilbert space as the shell of a tortoise, each invariant
subspace being one of the scales. Motion among different scales is
impossible. (See Fig.\ \ref{fig:tortoise} in the following.)

If $\mathrm{Tr} P_n=s_n<\infty$, then the limiting evolution
operator $\cV_n(t)$ (\ref{eq:slims}) within the subspace
$\cH_{P_n}$ has the form (\ref{eq:cVfin1}),
\beq
\cV_n (t)=P_n \exp(-i P_n H P_n t)  \label{eq:cVfin} .
\eeq
If $\cH_{P_n}\subset D (H)$, then the resulting Hamiltonian $P_n H
P_n$ is self-adjoint and $\cV_n(t)$ is unitary in $\cH_{P_n}$.

The original limiting result (\ref{eq:probinfu1}) is reobtained
when $p_n(0)=1$ for some $n$, in (\ref{eq:probinfu}): the initial
state is then in one of the invariant subspaces and the survival
probability in that subspace remains unity. However, even if the
limits are the same, notice that the setup described here is
conceptually different from that of Sec.\ \ref{sec-msth}. Indeed,
the dynamics (\ref{eq:boo}) allows transitions among different
subspaces $\cH_{P_n}\to\cH_{P_m}$, while the dynamics
(\ref{eq:Nproie}) completely forbids them. Therefore, for
\textit{finite} $N$, (\ref{eq:boo}) takes into account the
possibility that a given subspace $\cH_{P_n}$ gets repopulated
\cite{Nakazato96b,zenoreview} after the system has made
transitions to other subspaces, while in (\ref{eq:Nproie}) the
system must be found in $\cH_{P_n}$ at every measurement.

\section{Continuous observation}
\label{sec-QZEcont}
\andy{QZEcont}

The formulation of the preceding sections hinges upon von
Neumann's concept of ``projection" \cite{von}. A projection is
(supposed to be) an \emph{instantaneous} process, yielding the
``collapse" of the wave function, whose physical meaning has been
debated since the very birth of quantum mechanics \cite{NPN}.
Repeated projections in rapid succession yield the Zeno effect, as
we have seen.

A projection \emph{\`a la} von Neumann is a handy way to
``summarize" the complicated physical processes that take place
during a quantum measurement. A measurement process is performed
by an external (macroscopic) apparatus and involves dissipative
effects, that imply an interaction and an exchange of energy with
and often a flow of probability towards the environment. The
external system performing the observation need not be a
\emph{bona fide} detection system, namely a system that ``clicks"
or is endowed with a pointer. It is enough that the information on
the state of the observed system be encoded in the state of the
apparatus. For instance, a spontaneous emission process is often a
very effective measurement process, for it is irreversible and
leads to an entanglement of the state of the system (the emitting
atom or molecule) with the state of the apparatus (the
electromagnetic field). The von Neumann rules arise when one
traces away the photonic state and is left with an incoherent
superposition of atomic states. However, it is clear that the main
features of the Zeno effects would still be present if one would
formulate the measurement process in more realistic terms,
introducing a physical apparatus, a Hamiltonian and a suitable
interaction with the system undergoing the measurement. Such a
point of view was fully undertaken in \cite{zenoreview}, where a
novel and more general definition of QZE and IZE was given, that
makes no explicit use of projections \emph{\`{a} la} von Neumann.
It goes without saying that one can still make use of projection
operators, if such a description turns out to be simpler and more
economic (Occam's razor). However, a formulation of the Zeno
effects in terms of a Hamiltonian description is a significant
conceptual step. When such a formulation is possible and when the
Hamiltonian has (at most) a smooth dependence on time, we will
speak of QZE (or IZE) realized by means of a
\textit{continuous} measurement process.

A few examples will help us clarify these concept.

\subsection{Non-Hermitian Hamiltonian}
\label{sec-QZEnH}
\andy{QZEnH}

The effect of an external apparatus can be mimicked by a
non-Hermitian Hamiltonian. Consider a two-level system
\andy{+-2}
\beq
\langle 1| = (1,0), \quad
\langle 2| = (0,1),
\label{eq:+-2}
\eeq
with Hamiltonian
\andy{eq:nonhermham}
\beq
H _K= \pmatrix{0 & \Omega \cr \Omega & -i2K}
=\Omega(\ket{1}\bra{2}+\ket{2}\bra{1})-i2K \ket{2}\bra{2}.
\label{eq:nonhermham}
\eeq
This yields Rabi oscillations of frequency $\Omega$, but at the
same time absorbs away the $|2\rangle$ component of the Hilbert
space, performing in this way a ``measurement." Due to the
non-Hermitian features of this description, probabilities are not
conserved.

\begin{figure}[t]
\begin{center}
\includegraphics[width=11cm]{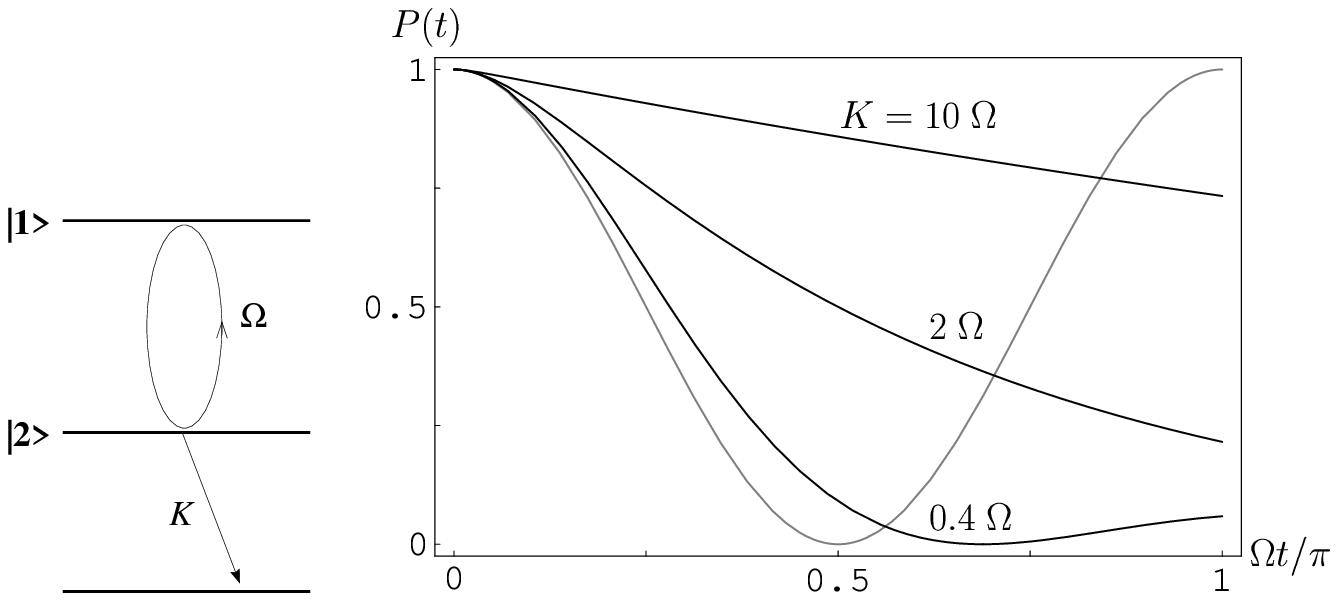}
\end{center}
\caption{Survival probability for a system undergoing Rabi
oscillations in presence of absorption ($K=0.4, 2, 10 \Omega$).
The gray line is the undisturbed evolution ($K=0$).}
\label{fig:zenoiV}
\end{figure}

Prepare the system in the initial state $\ket{1}$. An elementary
calculation \cite{zenoreview} yields the survival probability
\andy{survamplV}
\barr
p^{(K)}(t) = \left| \bra{1} e^{-i H_K t} \ket{1}\right|^2 &=&
\left| \frac{1}{2} \left( 1 + \frac{K}{\sqrt{K^2-\Omega^2}}
\right) e^{-(K-\sqrt{K^2-\Omega^2})t} \right.
\nonumber\\
& & + \left. \frac{1}{2} \left( 1 - \frac{K}{\sqrt{K^2-\Omega^2}}
\right) e^{-(K+\sqrt{K^2-\Omega^2})t} \right|^2, \;\;
\label{eq:survamplV}
\earr
which is shown in Fig.\
\ref{fig:zenoiV} for $K=0.4,2,10 \Omega$.
As expected, probability is (exponentially) absorbed away as $t
\to \infty$. However, as $K$ increases, the survival probability
reads
\andy{largeV}
\beq
p^{(K)}(t)\sim\left(1+\frac{\Omega^2}{2K}\right)
\exp\left(-\frac{\Omega^2}{K} t \right),
\qquad (t \gtrsim K^{-1})
\label{eq:largeV}
\eeq
and the effective decay rate $\gamma_{\mathrm{eff}}(K)=\Omega^2/K$
becomes smaller, eventually halting the ``decay" (and consequent
absorption) of the initial state and yielding an interesting
example of QZE: a larger $K$ entails a more ``effective"
measurement of the initial state. Notice that the expansion
(\ref{eq:largeV}) is not valid at very short times (where there is
a quadratic Zeno region), but becomes valid very quickly, on a
time scale of order $K^{-1}$ (the duration of the Zeno region
\cite{zenoreview,hydrogen,Antoniou}).

The (non-Hermitian) Hamiltonian (\ref{eq:nonhermham}) can be
obtained by considering the evolution engendered by a Hermitian
Hamiltonian acting on a larger Hilbert space and then restricting
the attention to the subspace spanned by $\{\ket{1}, \ket{2}\}$:
consider the Hamiltonian
\andy{hamflat}
\beq
\tilde H_K= \Omega(\ket{1}\bra{2}+\ket{2}\bra{1}) +\int
d\omega\;\omega \ket{\omega} \bra{\omega}
+\sqrt{\frac{2K}{\pi}}\int d\omega\;( \ket{2} \bra{\omega}+
\ket{\omega} \bra{2}) ,
\label{eq:hamflat}
\eeq
which describes a two-level system coupled to the photon field
$\{\ket{\omega}\}$ in the rotating-wave approximation. It is not
difficult to show \cite{zenoreview} that, if only state $\ket{1}$
is initially populated, this Hamiltonian is ``equivalent" to
(\ref{eq:nonhermham}), in that they both yield the \textit{same}
equations of motion in the subspace spanned by $\ket{1}$ and
$\ket{2}$. QZE is obtained by increasing $K$: a larger coupling to
the environment leads to a more effective ``continuous"
observation on the system (quicker response of the apparatus), and
as a consequence to a slower decay (QZE). The quantity $1/K$ is
the response time of the ``apparatus."

\subsection{Continuous Rabi observation}
\label{sec-QZEcont1}
\andy{QZEcont1}

The previous example might lead one to think that absorption
and/or probability leakage to the environment (or in general to
other degrees of freedom) are fundamental requisites to obtain
QZE. This expectation would be incorrect. Even more,
\textit{irreversibility is not essential}. Consider, indeed,
the 3-level system
\andy{levelM}
\beq
\langle 1| = (1,0,0), \quad
\langle 2| = (0,1,0), \quad
\langle 3| = (0,0,1)
\label{eq:levelM}
\eeq
and the (Hermitian) Hamiltonian
\andy{ham3l0}
\beq
H_{\mathrm{3lev}} = \Omega ( |1\rangle \langle 2| + |2\rangle
\langle 1|) + K ( |2\rangle \langle 3| + |3\rangle \langle 2|) =
\pmatrix{0 & \Omega & 0\cr \Omega & 0 & K \cr 0 & K & 0},
\label{eq:ham3l0}
\eeq
where $K \in \mathbb{R}$ is the strength of the coupling between
level $|2\rangle$ (``decay products") and level $3$ (that will
play the role of measuring apparatus). This model, first
considered by Peres \cite{Peres80}, is probably the simplest way
to include an ``external" apparatus in our description: as soon as
the system is in $|2\rangle$ it undergoes Rabi oscillations to
$|3\rangle$. We expect level $|3\rangle$  to perform better as a
measuring apparatus when the strength $K$ of the coupling becomes
larger.

\begin{figure}[t]
\begin{center}
\includegraphics[width=11cm]{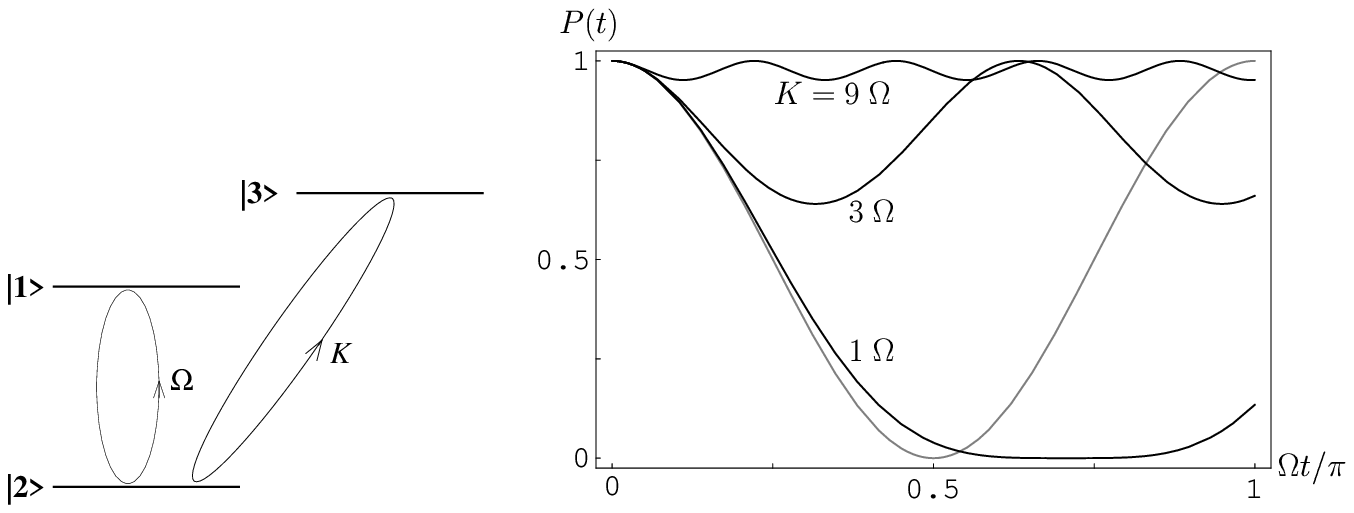}
\end{center}
\caption{Survival probability for a continuous Rabi ``measurement"
with $K=1,3,9\Omega$: quantum Zeno effect. The gray line is the
undisturbed evolution ($K=0$).}
\label{fig:zenocont}
\end{figure}

A straightforward calculation \cite{zenoreview} yields the
survival probability in the initial state $|1\rangle$
\andy{sp3}
\beq
p^{(K)}(t) = \left| \bra{1} e^{-i H_{\mathrm{3lev}} t}
\ket{1}\right|^2= \frac{1}{(K^2+\Omega^2)^2}\left[K^2+ \Omega^2
\cos(\sqrt{K^2+\Omega^2}t) \right]^2. \label{eq:sp3}
\eeq
This is shown in Fig.\ \ref{fig:zenocont} for $K=1,3,9 \Omega$.
We notice that for large $K$ the state of the system does not
change much: as $K$ is increased, level $|3\rangle$ performs a
better ``observation" of the state of the system, hindering
transitions from $|1\rangle$ to $|2\rangle$. This can be viewed as
a QZE due to a ``continuous," yet Hermitian observation performed
by level $|3\rangle$.

In spite of their simplicity, the models shown in this section
clarify the physical meaning of a ``continuous" measurement
performed by an ``external apparatus" (which can even be another
degree of freedom of the system investigated). Also, they capture
and elucidate many interesting features of a Zeno dynamics.

\section{Novel definition of quantum Zeno effect}
\label{sec-novdef}
\andy{novdef}

The examples considered in the previous section call for a broader
formulation of Zeno effect, that should be able to include
``continuous" observations as well as other situations that do not
fit into the scheme of the ``pulsed" formulation. We proposed such
a definition in Ref.\ \cite{zenoreview}. It comprises all possible
cases (oscillating as well as unstable systems) and situations
(quantum Zeno effect as well as inverse quantum Zeno effect).
Although in this article we are mostly concerned with the QZE for
oscillating systems, we give here all definitions for the sake of
completeness.

Consider a quantum system whose evolution is described by a
Hamiltonian $H$. Let the initial state be $\rho_0$ (not
necessarily a pure state) and its survival probability $p(t)$.
Consider the evolution of the system under the effect of an
additional interaction, so that the total Hamiltonian reads
\andy{HK}
\beq
H_K= H + H_{\mathrm{meas}}(K),
 \label{eq:HK}
\eeq
where $K$ is a set of parameters (such as coupling constants) and
$H_{\mathrm{meas}}(K=0)=0$. Notice that $H$ is not necessarily the
free Hamiltonian; rather, one should think of $H$ as a full
Hamiltonian, containing interaction terms, and
$H_{\mathrm{meas}}(K)$ should be viewed as an ``additional"
interaction Hamiltonian performing the ``measurement." If $K$ is
simply a coupling constant, then the above formula simplifies to
\andy{HKcoup}
\beq
H_K= H + K H_{\mathrm{meas}} .
 \label{eq:HKcoup}
\eeq
Notice that if a projection is viewed as a shorthand notation for
a (generalized \cite{PN94}) spectral decomposition
\cite{Wigner63}, the above Hamiltonian scheme includes, for all
practical purposes, the usual formulation of quantum Zeno effect
in terms of projection operators. In such a case the scheme
(\ref{eq:HK}) is more appropriate, for a fine tuning of $K$ might
be required \cite{PN94}.

All the examples considered in the previous sections (for both
``pulsed" and ``continuous" measurements) can be analyzed within
the scheme (\ref{eq:HKcoup}) and \emph{a fortiori} (\ref{eq:HK}).
We can now define all possible Zeno effects.

\subsection{Oscillating systems}
\label{sec-novdefosc}
\andy{novdefosc}

We shall say that an oscillating system displays a QZE if there
exist an interval $I^{(K)}=[t_1^{(K)},t_2^{(K)}]$ such that
\andy{defqze}
\beq
p^{(K)}(t) > p(t), \qquad \forall t \in I^{(K)},
 \label{eq:defqze}
\eeq
where $p^{(K)}(t)$ and $p(t)=p^{(0)}(t)$ are the survival
probabilities under the action of the Hamiltonians $H_K$ and $H$,
respectively. We shall say that the system displays an IZE if
there exist an interval $I^{(K)}$ such that
\andy{defiqze}
\beq
p^{(K)}(t) < p(t), \qquad \forall t \in I^{(K)} .
 \label{eq:defiqze}
\eeq
The time interval $I^{(K)}$ must be evaluated case by case.
However,
\andy{t2lim}
\beq
t_2^{(K)} \leq T_{\mathrm{P}},
 \label{eq:t2lim}
\eeq
where $T_{\mathrm{P}}$ is the Poincar\'e time of the system.
Obviously, in order that the definition
(\ref{eq:defqze})-(\ref{eq:defiqze}) be meaningful from a physical
point of view, the length of the interval $I^{(K)}$ must be of
order $T_{\mathrm{P}}$.

The above definition is very broad and includes a huge class of
systems [even trivial cases such as time translations $p(t) \to
p(t-t_0)$]. We would like to stress that we have not succeeded in
finding a more restrictive definition and we do not think it would
be meaningful: many phenomena can be viewed or reinterpreted as
Zeno effects and this is in our opinion a fecund point of view
\cite{zenoreview}.

In order to elucidate the meaning of the above definition, let us
look at some particular cases considered in the previous sections.
The situations considered in Figs.\ \ref{fig:zenoiV} and
\ref{fig:zenocont} are both QZEs, according to this definition:
one has $t_1^{(K)}=0$ and $t_2^{(K)} \leq
T_{\mathrm{P}}=\pi/\Omega$ [and $(t_2^{(K)}-t_1^{(K)}) = \Ord
(T_{\mathrm{P}}$)]. The case outlined in Fig.\ \ref{fig:zenoevol}
is also a QZE, with $t_1^{(K)}=0$ and $t_2^{(K)} \leq
T_{\mathrm{P}}$ (notice that $T_{\mathrm{P}}$ may even be
infinite).

\subsection{Unstable systems}
\label{sec-novdefunst}
\andy{novdefunst}

In this paper we mostly deal with few-level systems. However, for
unstable systems, the definition of Zeno effect can be made more
stringent and expressed in terms of a single parameter, the decay
rate. In fact, in such a case, one need not refer to a given
interval $I^{(K)}$, but can consider the \emph{global} behavior of
the survival probability.

Let us consider Eqs.\ (\ref{eq:Hdiv}) and (\ref{eq:HKcoup}). For
an unstable system, the off-diagonal interaction Hamiltonian
$H_{\mathrm{int}}$ in Eq.\ (\ref{eq:Hdiv}) is responsible for the
decay. Let
\beq
\label{eq:gammadef}
\gamma = 2 \pi \bra{a} H_{\mathrm{int}} \delta(\omega_a-H_0)
H_{\mathrm{int}} \ket{a}
\eeq
be the decay rate (Fermi ``golden" rule
\cite{Fermi}, valid at second order in the decay coupling
constant), $\ket{a}$ being the initial state, which is an
eigenstate of $H_0$ with energy $\omega_a$. We define the
occurrence of a QZE or an IZE if
\andy{defgammaze}
\beq
\gamma_{\mathrm{eff}}(K) \;^<_> \; \gamma,
 \label{eq:defgammaze}
\eeq
respectively, where $\gamma_{\mathrm{eff}}(K)$ is the new
(effective) decay rate under the action of $H_K$,
\beq
\label{eq:gammaKdef}
\gamma_{\mathrm{eff}}(K) = 2 \pi \bra{a} (H_{\mathrm{int}}+ K
H_{\mathrm{meas}})\, \delta(\omega_a-H_0)\, (H_{\mathrm{int}}+ K
H_{\mathrm{meas}}) \ket{a}.
\eeq
Notice that this case is in agreement with the definitions\
(\ref{eq:defqze})-(\ref{eq:defiqze}). Moreover, $t_2^{(K)} \to
\infty$ for IZE, while $t_2^{(K)} \leq t_{\mathrm{pow}}$ for QZE,
where $t_{\mathrm{pow}}$ is the time at which a transition from an
exponential to a power law takes place. (Such a time is of order
$\log$(coupling constant), at least for renormalizable quantum
field theories \cite{vanhove}.)

It is worth noticing that (\ref{eq:defgammaze}) is of general
validity when it refers to physical decay rates, even when the
perturbative expressions (\ref{eq:gammadef}) and
(\ref{eq:gammaKdef}) are not valid. In such a case the decay rate
is simply given by the imaginary part of the pole
$E_{\mathrm{pole}}$ of the resolvent nearest to the real axis in
the second Riemann sheet of the complex energy plane
\cite{strev}. The pole is the solution of the equation
\beq
E_{\mathrm{pole}}= \omega_a + \Sigma_{\mathrm{II}}
(E_{\mathrm{pole}}), \qquad \gamma=-2 \Im [E_{\mathrm{pole}} ],
\eeq
where $\Sigma_{\mathrm{II}} (E)$ is the determination of the
proper self-energy function
\beq
\Sigma (E)=\bra{a} H_{\mathrm{int}} \frac{1}{E-H_0}
H_{\mathrm{int}} \ket{a}
\label{eq:selfenergy}
\eeq
on the second Riemann sheet. Analogously for
$\gamma_{\mathrm{eff}}(K)$, with the substitution
$H_{\mathrm{int}}\to H_{\mathrm{int}}+ K H_{\mathrm{meas}}$ in
Eq.\ (\ref{eq:selfenergy}). For a more detailed discussion, see
\cite{zenoreview}.

\section{Dynamical quantum Zeno effect}
 \label{sec-dynamicalQZE}
 \andy{dynamicalQZE}

The broader formulation of quantum Zeno effect (and inverse
quantum Zeno effect) elaborated in Sec.\ \ref{sec-novdef} triggers
a spontaneous question about the form of the interaction
Hamiltonian $H_{\mathrm{meas}}$ between system and apparatus [Eq.\
(\ref{eq:HKcoup})]. In the case of pulsed measurements, in order
to get a Zeno effect one has to prepare the system in a state
belonging to the measured subspace $\cH_P$ as in Eq.\
(\ref{eq:inprep}) [or to any subspace $\cH_{P_n}$ of the partition
(\ref{eq:partition}) for nonselective measurements]. On the other
hand, in the case of a continuous measurement it is not clear
which relation must hold between the initial state of the system
$\rho_0$ and the structure of the interaction Hamiltonian
$H_{\mathrm{meas}}$ in order to get a Zeno effect. We have
introduced two paradigmatic examples in Sec.\ \ref{sec-QZEcont},
but we still do not know \textit{why} they work. It is therefore
important to understand in more details which features of the
coupling between the ``observed" system and the ``measuring"
apparatus are needed to obtain a QZE. In other words, one wants to
know when an external quantum system can be considered a good
apparatus and why. We shall try to clarify these issues and cast
the dynamical quantum Zeno evolution in terms of an adiabatic
theorem. We will show that the evolution of a quantum system under
the action of a continuous measurement process is in fact similar
to that obtained with pulsed measurements: the system is forced to
evolve in a set of orthogonal subspaces of the total Hilbert space
and an effective superselection rule arises in the strong coupling
limit. These \textit{quantum Zeno subspaces} \cite{theorem} are
just the eigenspaces (belonging to different eigenvalues) of the
Hamiltonian describing the interaction between the system and the
apparatus: they are subspaces that the measurement process is able
to distinguish.

\subsection{A theorem}
 \label{sec-dyntheo}
 \andy{dyntheo}

Our answer to the afore-mentioned question is contained in a
theorem
\cite{thesis,theorem}, which is the exact analog of Misra and
Sudarshan's theorem for a general dynamical evolution of the type
(\ref{eq:HKcoup}). Consider the time evolution operator
\barr
U_{K}(t) = \exp(-iH_K t) . \label{eq:measinter}
\earr
We will prove that in the ``infinitely strong measurement"
(``infinitely quick detector") limit $K\to\infty$ the evolution
operator
\beq
\label{eq:limevol} \cU(t)=\lim_{K\to\infty}U_{K}(t),
\eeq
becomes diagonal with respect to $H_{\mathrm{meas}}$:
\beq
\label{eq:diagevol} [\cU(t), P_n]=0, \qquad \mathrm{where} \quad
H_{\mathrm{meas}} P_n=\eta_n P_n,
\eeq
$P_n$ being the orthogonal projection onto $\cH_{P_n}$, the
eigenspace of $H_{\mathrm{meas}}$ belonging to the eigenvalue
$\eta_n$. Note that in Eq.\ (\ref{eq:diagevol}) one has to
consider distinct eigenvalues, i.e., $\eta_n\neq\eta_m$ for $n\neq
m$, whence the $\cH_{P_n}$'s are in general multidimensional.

Moreover, the limiting evolution operator has the explicit form
\beq
\label{eq:theorem} \cU(t)=\exp[-i(H_{\mathrm{diag}}+K
H_{\mathrm{meas}}) t],
\eeq
where
\beq
H_{\mathrm{diag}}=\sum_n P_n H P_n \label{eq:diagsys}
\eeq
is the diagonal part of the system Hamiltonian $H$ with respect to
the interaction Hamiltonian $H_{\mathrm{meas}}$.

In conclusion, the generator of the dynamics is the \textit{Zeno
Hamiltonian}
\andy{Zenogenerator}
\beq
H^{\mathrm{Z}} = H_{\mathrm{diag}}+K H_{\mathrm{meas}} = \sum_n
\left( P_n H P_n + K \eta_n P_n \right), \label{eq:Zenogenerator}
\eeq
whose diagonal structure is explicit, and the evolution operator
is
\andy{eq:measinterbis}
\barr
\cU(t) = \exp(-iH^{\mathrm{Z}} t) . \label{eq:measinterbis}
\earr

\subsection{Dynamical superselection rules}
\label{sec-supersel}
\andy{supersel}

Before proving the theorem of Sec.\ \ref{sec-dyntheo} let us
briefly consider its physical implications. In the $K\to\infty$
limit, due to (\ref{eq:diagevol}), the time evolution operator
becomes diagonal with respect to $H_{\mathrm{meas}}$,
\beq
[\cU(t), H_{\mathrm{meas}}]=0,
\eeq
a superselection rule arises and the total Hilbert space is split
into subspaces $\cH_{P_n}$ which are invariant under the
evolution. These subspaces are simply defined by the $P_n$'s,
i.e., they are eigenspaces belonging to distinct eigenvalues
$\eta_n$: in other words, they are \textit{subspaces that the
apparatus is able to distinguish}. On the other hand, due to
(\ref{eq:Zenogenerator})-(\ref{eq:measinterbis}), the dynamics
within each Zeno subspace $\cH_{P_n}$ is essentially governed by
the diagonal part $P_n H P_n$ of the system Hamiltonian $H$ (the
remaining part of the evolution consisting in a (sector-dependent)
phase). The evolution reads
\barr
\label{eq:dynamicalQZE}
\rho(t)= \cU(t) \rho_0 \cU^\dagger(t) =e^{-i H^{\mathrm{Z}}
t}\rho_0 e^{i H^{\mathrm{Z}} t}
\earr
and the probability to find the system in each $\cH_{P_n}$
\barr
p_n(t)&=&\mathrm{Tr} \left[ \rho(t) P_n \right]= \mathrm{Tr}
\left[\cU(t)\rho_0\cU^\dagger(t) P_n\right] =\mathrm{Tr}
\left[\cU(t)\rho_0 P_n\cU^\dagger(t)\right]
\nonumber\\
&=& \mathrm{Tr} \left[ \rho_0 P_n \right]=p_n(0)
\label{eq:pntpn0}
\earr
is constant. As a consequence, if the initial state of the system
belongs to a specific sector, it will be forced to remain there
forever (QZE):
\beq
\psi_0\in \cH_{P_n}\rightarrow \psi(t)\in \cH_{P_n}.
\eeq
\begin{figure}[t]
\begin{center}
\includegraphics[height=6.5cm]{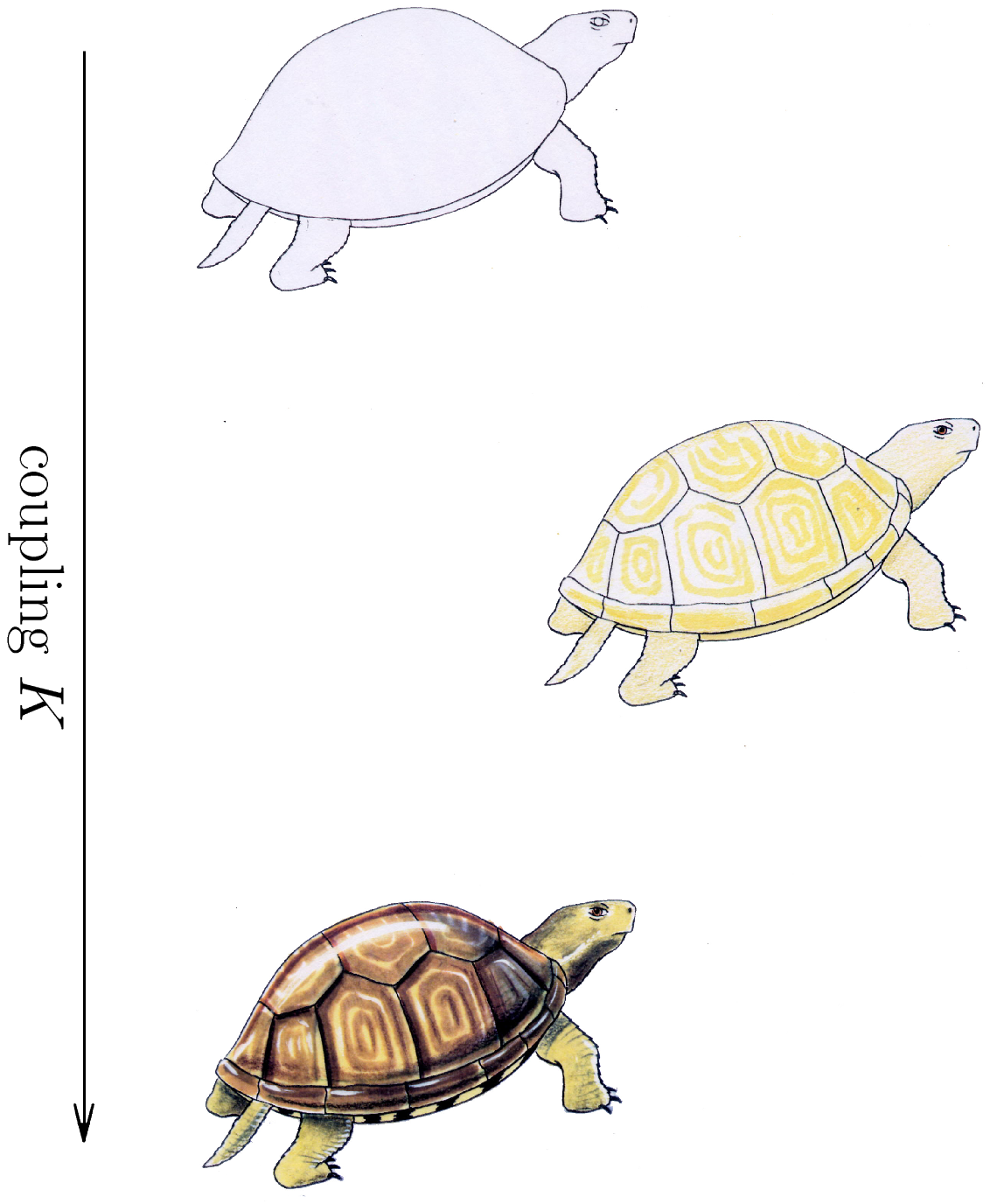}
\end{center}
\caption{\label{fig:tortoise} The Hilbert space of the system: a
dynamical superselection rule appears as the coupling $K$ to the
apparatus is increased.}
\end{figure}
More generally, if the initial state is an incoherent
superposition of the form $\rho_0=\hat P \rho_0$, with $\hat P$
defined in (\ref{eq:superP}), then each component will evolve
separately, according to
\barr
\rho(t)&=&\cU(t)\rho_0\cU^\dagger(t)=\sum_n e^{-i H^{\mathrm{Z}}t}
P_n\rho_0 P_n e^{i H^{\mathrm{Z}} t}
\nonumber\\
&=&\sum_n e^{-i P_n H P_n t} P_n\rho_0 P_n e^{i P_n H P_n t}=
\sum_n \cV_n(t) \rho_0 \cV_n^\dagger(t),
\label{eq:cUcV}
\earr
with $\cV_n(t)= P_n \exp(-i P_n H P_n t)$, which is exactly the
same result (\ref{eq:rhoZ})-(\ref{eq:cVfin}) found in the case of
nonselective pulsed measurements.  This bridges the gap with the
description of Sec.\ \ref{sec-nonselect} and clarifies the role of
the detection apparatus: it defines the Zeno subspaces. In Fig.\
\ref{fig:tortoise} we endeavored to give a pictorial representation of
the decomposition of the Hilbert space as $K$ is increased.

Notice, however, that there is one important difference between
the dynamical evolution (\ref{eq:dynamicalQZE}) and the projected
evolution (\ref{eq:rhoZ}). Indeed, if the initial state $\rho_0$
contains coherent terms between any two Zeno subspaces $\cH_{P_n}$
and $\cH_{P_m}$, $P_n \rho_0 P_m\neq 0$, these vanish after the
first projection in (\ref{eq:rhoZ}), $P_n
\rho(0^+) P_m = 0$, and the state becomes an
incoherent superposition $\rho(0^+)\neq\rho_0$, whence
$\mathrm{Tr}
\rho(0^+)^2 < \mathrm{Tr} \rho_0^2$. On the other hand, such terms
are preserved by the dynamical (unitary) evolution
(\ref{eq:dynamicalQZE}) and do not vanish, even though they wildly
oscillate. For example, consider the initial state
\beq
\rho_0= (P_n+P_m)\rho_0 (P_n+P_m), \qquad P_n \rho_0 P_m \neq 0.
\eeq
By Eq.\ (\ref{eq:dynamicalQZE}) it evolves into
\barr
\rho(t)&=&\cV_n (t) \rho_0 \cV^\dagger_n(t)+ \cV_m (t) \rho_0
\cV^\dagger_m(t)
\nonumber\\
& & + e^{-i K (\eta_n-\eta_m) t} \cV_n (t) \rho_0 \cV^\dagger_m(t)
+e^{i K (\eta_n-\eta_m) t} \cV_m (t) \rho_0 \cV^\dagger_n(t) ,
\earr
at variance with (\ref{eq:cUcV}) and (\ref{eq:rhoZ}). Therefore
$\mathrm{Tr} \rho(t)^2 = \mathrm{Tr} \rho_0^2$ for any $t$ and the
Zeno dynamics is unitary in the \textit{whole} Hilbert space
$\cH$. We notice that these coherent terms become unobservable in
the large-$K$ limit, as a consequence of the Riemann-Lebesgue
theorem (applied to any observable that ``connects" different
sectors and whose time resolution is finite). This interesting
aspect is reminiscent of some results on ``classical" observables
\cite{Jauch}, semiclassical limit \cite{Berry} and quantum
measurement theory \cite{Araki,Schwinger59}.

It is worth noticing that the superselection rules discussed here
are \textit{de facto} equivalent to the celebrated ``W$^3$" ones
\cite{WWW}, but turn out to be a mere consequence of the Zeno
dynamics. For a related discussion, but in a different context,
see \cite{Giulini}.

\subsection{Proof of the theorem}
 \label{sec-proof}
 \andy{proof}

We will now use perturbation theory and prove \cite{thesis} that
the limiting evolution operator has the form (\ref{eq:theorem}).
Property (\ref{eq:diagevol}) will then automatically follow. In
the next subsection we will give a more direct proof of
(\ref{eq:diagevol}), which relies on the adiabatic theorem.

Rewrite the time evolution operator in the form
\beq
\label{eq:timevoltau} U_K(t)=\exp(-i H_K t)=\exp(-i H_\lambda
\tau)=U_\lambda(\tau)
\eeq
where
\beq
\lambda=1/K,\qquad \tau= K t=t/\lambda, \qquad H_\lambda=\lambda
H_K= H_{\mathrm{meas}}+\lambda H, \label{eq:lambdaK}
\eeq
and apply  perturbation theory to the Hamiltonian $H_\lambda$ for
small $\lambda$. To this end, choose the unperturbed degenerate
projections $P_{n\alpha}$
\beq
H_{\mathrm{meas}}P_{n\alpha}=\eta_n P_{n\alpha}, \qquad P_n=
\sum_{\alpha}P_{n\alpha},
\eeq
whose degeneration $\alpha$ is resolved at some order in the
coupling constant $\lambda$. This means that by denoting
$\widetilde \eta_{n\alpha}$ and $\widetilde P_{n\alpha}$ the
eigenvalues and the orthogonal projections of the total
Hamiltonian $H_{\lambda}$,
\beq
H_\lambda\widetilde{P}_{n\alpha}=\widetilde{\eta}_{n\alpha}\widetilde{P}_{n\alpha},
\eeq
they reduce to the unperturbed ones when the perturbation vanishes
\beq
\widetilde{P}_{n\alpha}\stackrel{\lambda\to
0}{\longrightarrow}P_{n\alpha},\qquad
\widetilde{\eta}_{n\alpha}\stackrel{\lambda\to
0}{\longrightarrow}\eta_n.
\eeq
Therefore, by applying standard perturbation theory
\cite{Messiah61}, we get the eigenprojections
\barr
\widetilde{P}_{n\alpha}&=&P_{n\alpha}+\lambda
P_{n\alpha}^{(1)}+\Ord(\lambda^2)
\nonumber\\
&=&P_{n\alpha}+\lambda \left( \frac{Q_n}{a_n}H
P_{n\alpha}+P_{n\alpha}H\frac{Q_n}{a_n}\right)+\Ord(\lambda^2) ,
\label{eq:perteigenvec}
\earr
where
\barr
Q_n=1-P_n=\sum_{m\neq n}P_{m}, \qquad
\frac{Q_n}{a_n}=\frac{Q_n}{\eta_n-H_{\mathrm{meas}}}=\sum_{m\neq
n}\frac{P_m}{\eta_n-\eta_m} .
\earr
The perturbative expansion of the eigenvalues reads
\barr
\widetilde{\eta}_{n\alpha}=\eta_n+\lambda
\eta_{n\alpha}^{(1)}+\lambda^2
\eta_{n\alpha}^{(2)}+\Ord(\lambda^3) \label{eq:perteigenval}
\earr
where
\barr
\eta_{n\alpha}^{(1)} P_{n\alpha} &=& P_{n\alpha} H P_{n\alpha},
\qquad
\eta_{n\alpha}^{(2)} P_{n\alpha} = P_{n\alpha} H \frac{Q_n}{a_n} H
P_{n\alpha},
\nonumber\\
P_{n\alpha} H P_{n\beta} &=& P_{n\alpha} H \frac{Q_n}{a_n} H
P_{n\beta}=0,\qquad \alpha\neq\beta . \label{eq:perteigenval1}
\earr
 Write now the spectral decomposition of the evolution operator\
(\ref{eq:timevoltau}) in terms of the projections $\widetilde
P_{n\alpha}$
\beq
U_\lambda(\tau)=\exp(-iH_\lambda
\tau)\sum_{n,\alpha}\widetilde{P}_{n\alpha}=
\sum_{n,\alpha}\exp(-i\widetilde{\eta}_{n\alpha}
\tau)\widetilde{P}_{n\alpha}
\eeq
and plug in the perturbation expansions\ (\ref{eq:perteigenvec}),
to obtain
\barr
U_\lambda(\tau)
&=&\sum_{n,\alpha}e^{-i\widetilde{\eta}_{n\alpha}\tau}P_{n\alpha}
\nonumber\\
& & +\lambda\sum_{n,\alpha}\left(\frac{Q_n}{a_n}H P_{n\alpha}
e^{-i\widetilde{\eta}_{n\alpha}\tau}
+e^{-i\widetilde{\eta}_{n\alpha}\tau}P_{n\alpha}H\frac{Q_n}{a_n}
\right)+\Ord(\lambda^2).
\label{eq:Upertexp}
\earr
Let us define the operator
\barr
\widetilde{H}_{\lambda}&=&\sum_{n,\alpha}\widetilde{\eta}_{n\alpha}P_{n\alpha}
\nonumber\\
&=&H_{\mathrm{meas}}+\lambda\sum_n P_n H P_n+ \lambda^2\sum_n P_n
H \frac{Q_n}{a_n}H P_n+\Ord(\lambda^3),
\label{eq:tildeHl}
\earr
where Eqs.\ (\ref{eq:perteigenval})-(\ref{eq:perteigenval1}) were
used. By plugging Eq.\ (\ref{eq:tildeHl}) into Eq.\
(\ref{eq:Upertexp}) and making use of the property
\beq
\sum_n P_n H \frac{Q_n}{a_n}=-\sum_n\frac{Q_n}{a_n}H P_{n} ,
\eeq
we finally obtain
\barr
U_\lambda(\tau) =\exp(-i\widetilde{H}_\lambda \tau)+ \lambda\left[
\sum_n\frac{Q_n}{a_n}H P_{n}, \; \exp(-i\widetilde{H}_\lambda
\tau)\right] +\Ord(\lambda^2). \label{eq:Upertexp1}
\earr
Now, by recalling the definition\ (\ref{eq:lambdaK}), we can write
the time evolution operator $U_K(t)$ as the sum of two terms
\andy{eq:adK}
\beq
\label{eq:adK}
U_K(t)=U_{\mathrm{ad}, K}(t)+\frac{1}{K} U_{\mathrm{na},K}(t),
\eeq
where
\barr
U_{\mathrm{ad},K}(t)=e^{-i\left( K H_{\mathrm{meas}}+\sum_n P_n H
P_n+
\frac{1}{K}\sum_n P_n H \frac{Q_n}{a_n}H
P_n+\Ord\left(K^{-2}\right)\right)t} \;\;\;\;
\earr
is a diagonal, \textit{adiabatic} evolution and
\beq
U_{\mathrm{na},K}(t)=\left[\sum_n
\frac{Q_n}{a_n}H P_n,\; U_{\mathrm{ad},K}(t)\right]+\Ord\left(K^{-1}\right)
\eeq
is the off-diagonal, \textit{nonadiabatic} correction. In the
$K\to\infty$ limit only the adiabatic term survives and one
obtains
\beq
\cU(t)=\lim_{K\to\infty}U_{K}(t)=\lim_{K\to\infty}
U_{\mathrm{ad},K}(t)=e^{-i \left(K H_{\mathrm{meas}}+\sum_n P_n H
P_n\right) t},
\label{eq:theorem2}
\eeq
which is formula\ (\ref{eq:theorem}) [and implies also
(\ref{eq:diagevol})]. The proof is complete. As a byproduct we get
the corrections to the exact limit, valid for large, but finite,
values of $K$.

Notice that in our derivation we assumed that the eigenprojections
and the eigenvalues of the perturbed Hamiltonian $H_{\lambda}$
admit the asymptotic expansions (\ref{eq:perteigenvec}) and
(\ref{eq:perteigenval}) up to order $\Ord(\lambda^2)$ and
$\Ord(\lambda^3)$, respectively. With these assumptions we have
been able to exhibit also the first corrections to the limit.
However, it is apparent that in order to prove the limit
(\ref{eq:theorem2}), it is sufficient to assume that the
eigenprojections and the eigenvalues admit the expansions
\barr
\widetilde{P}_{n\alpha}=P_{n\alpha}+\mathrm{o}(1),\qquad
\widetilde{\eta}_{n\alpha}=\eta_n+\lambda \eta_{n\alpha}^{(1)}+
\mathrm{o}(\lambda),\qquad \mathrm{for}\quad\lambda\to 0,
\label{eq:perteigenval3}
\earr
whence
\beq
U_{K}(t)=e^{-i \left[K H_{\mathrm{meas}}+\sum_n P_n H
P_n+\mathrm{o}(1)\right] t}+\mathrm{o}(1),\qquad \mathrm{for}\quad
K\to \infty.
\label{eq:theorem3}
\eeq
Notice however that in such a case, unlike in (\ref{eq:adK}), we
have no information on the approaching rate and the first-order
corrections.

\subsection{Zeno evolution from an adiabatic theorem}
 \label{sec-Zenoadiab}
 \andy{Zenoadiab}

We now give an alternative proof [and a generalization to
time-dependent Hamiltonians $H(t)$] of Eq.\ (\ref{eq:diagevol}).
We follow again \cite{thesis}. The adiabatic theorem deals with
the time evolution operator $U(t)$ when the Hamiltonian $H(t)$
slowly depends on time. The traditional formulation
\cite{Messiah61} replaces the physical time $t$ by the scaled time
$s=t/T$ and considers the solution of the scaled Schr\"odinger
equation
\beq
i \frac{d}{ds} U_T (s) = T H(s) U_T (s) \label{eq:adiabatic}
\eeq
in the $T\to\infty$ limit.

Given a family  $P(s)$ of smooth spectral projections of $H(s)$
\beq
H(s)P(s)=E(s)P(s),
\eeq
the adiabatic time evolution $U_{\mathrm{A}}(s)=\lim_{T\to\infty}
U_T(s)$ has the intertwining property \cite{adiabatic,Messiah61}
\beq
U_{\mathrm{A}}(s)P(0)=P(s) U_{\mathrm{A}} (s) ,
\label{eq:adintert}
\eeq
that is, $U_{\mathrm{A}}(s)$ maps $\cH_{P(0)}$ onto $\cH_{P(s)}$.

Theorem (\ref{eq:diagevol}) and its generalization,
\beq
\cU(t) P_n(0)=P_n(t) \cU(t), \label{eq:theoremt}
\eeq
valid for generic time dependent Hamiltonians,
\beq
\label{eq:sys+meast} H_K(t)=H(t)+ K H_{\mathrm{meas}}(t),
\eeq
are easily proven by recasting them in the form of an adiabatic
theorem \cite{theorem}. In the $H$ interaction picture, given by
\beq
i \frac{d}{dt} U_{\mathrm{S}}(t) = H U_{\mathrm{S}}(t), \qquad
H^{\mathrm{I}}_{\mathrm{meas}}(t)=U^\dagger_{\mathrm{S}}(t)
H_{\mathrm{meas}} U_{\mathrm{S}}(t),
\eeq
the Schr\"odinger equation reads
\beq
i \frac{d}{dt} U_K^{\mathrm{I}} (t) = K
H^{\mathrm{I}}_{\mathrm{meas}}(t) \; U_K^{\mathrm{I}} (t).
\label{eq:adiabaticlike}
\eeq
The Zeno evolution pertains to the $K\to\infty$ limit: in such a
limit Eq.\ (\ref{eq:adiabaticlike}) has exactly the same form of
the adiabatic evolution (\ref{eq:adiabatic}): the large coupling
$K$ limit corresponds to the large time $T$ limit and the physical
time $t$ to the scaled time $s=t/T$. Therefore, let us consider a
spectral projection of $H^{\mathrm{I}}_{\mathrm{meas}}(t)$,
\beq
P^{\mathrm{I}}_n(t)=U^\dagger_{\mathrm{S}}(t) P_n(t)
U_{\mathrm{S}}(t) , \label{eq:PIn}
\eeq
such that
\beq
H^{\mathrm{I}}_{\mathrm{meas}}(t) P^{\mathrm{I}}_n(t)= \eta_n(t)
P^{\mathrm{I}}_n(t), \qquad H_{\mathrm{meas}}(t) P_n(t)= \eta_n(t)
P_n(t) .
\eeq
The limiting operator
\beq
\cU^{\mathrm{I}} (t)=\lim_{K\to\infty}U_K^{\mathrm{I}}(t)
\label{eq:limitint}
\eeq
has the intertwining property (\ref{eq:adintert})
\beq
\cU^{\mathrm{I}} (t)
P^{\mathrm{I}}_n(0)=P^{\mathrm{I}}_n(t)\cU^{\mathrm{I}} (t),
\label{eq:intertwinint}
\eeq
i.e.\ maps $\cH_{P^{\mathrm{I}}_n(0)}$ onto
$\cH_{P^{\mathrm{I}}_n(t)}$:
\beq
\psi^{\mathrm{I}}_0\in \cH_{P^{\mathrm{I}}_n(0)}\rightarrow
\psi^{\mathrm{I}}(t)\in \cH_{P^{\mathrm{I}}_n(t)}.
\eeq
In the Schr\"odinger picture the limiting operator
\beq
\cU(t)=\lim_{K\to\infty} U_K(t)=\lim_{K\to\infty}
U_{\mathrm{S}}(t)
U_K^{\mathrm{I}}(t)=U_{\mathrm{S}}(t)\cU^{\mathrm{I}}(t)
\label{eq:limitSch}
\eeq
satisfies the intertwining property (\ref{eq:theoremt}) [see
(\ref{eq:PIn})]
\barr
\cU(t) P_n(0) &=& U_{\mathrm{S}}(t)\cU^{\mathrm{I}}(t) P_n(0)
=U_{\mathrm{S}}(t)\cU^{\mathrm{I}} (t)
P^{\mathrm{I}}_n(0)\nonumber\\
&=&U_{\mathrm{S}}(t) P^{\mathrm{I}}_n(t)\cU^{\mathrm{I}} (t) =
P_n(t) U_{\mathrm{S}}(t) \cU^{\mathrm{I}}(t)=P_n(t) \cU(t),
\label{eq:intertwin}
\earr
and maps $\cH_{P_n(0)}$ onto $\cH_{P_n(t)}$:
\beq
\psi_0\in \cH_{P_n(0)}\rightarrow \psi(t)\in \cH_{P_n(t)}.
\label{eq:mapS}
\eeq
The probability to find the system in $\cH_{P_n(t)}$,
\barr
p_n(t)&=&\mathrm{Tr}\left[P_n(t) \cU(t) \rho_0
\cU^\dagger(t)\right] =\mathrm{Tr}\left[\cU(t) P_n(0)\rho_0
\cU^\dagger(t)\right]\nonumber\\
&=&\mathrm{Tr}\left[P_n(0)\rho_0 \right]=p_n(0),
\label{eq:pntpn01}
\earr
is constant: if the initial state of the system belongs to a given
sector, it will be forced to remain there forever (QZE).

For a time-independent Hamiltonian
$H_{\mathrm{meas}}(t)=H_{\mathrm{meas}}$, the projections are
constant, $P_n(t)=P_n$, hence Eq.\ (\ref{eq:theoremt}) reduces to
(\ref{eq:diagevol}) and the above property holds
\textit{a fortiori} and reduces to (\ref{eq:pntpn0}).

Let us add a few comments. It is worth noticing that the limiting
evolutions (\ref{eq:limevol}), (\ref{eq:limitint}) and
(\ref{eq:limitSch}) are understood in the sense of the
intertwining relations (\ref{eq:diagevol}),
(\ref{eq:intertwinint}) and (\ref{eq:intertwin}), that is
\beq
\lim_{K\to \infty} \Big( U_K P_n - P_n U_K \Big) = 0 ,
\eeq
while, strictly speaking, each single addend has no limit, due to
a fast oscillating phase. In other words, one would read Eq.\
(\ref{eq:intertwin}) as
\barr
U_K(t) P_n(0)- P_n(t) U_K(t)=\mathrm{o}(1),
\qquad\mathrm{for}\quad K\to\infty.
\label{eq:theoremad}
\earr
As a matter of fact, there is no single adiabatic theorem
\cite{Avron}. Different adiabatic theorems follow from different
assumptions about the properties of
$H^{\mathrm{I}}_{\mathrm{meas}}(t)$ and $P^{\mathrm{I}}_n(t)$, the
notion of smoothness, what are the optimal error estimates, and so
on. But all these theorems have the structure of
Eq.~(\ref{eq:theoremad}) and only differ in their respective
approaching rates [for example, for noncrossing energy levels,
$\mathrm{o}(1)$ is in fact $\Ord(1/K)$, while for crossing levels
the rate is $\Ord(1/\sqrt{K})$]. The theorem we have shown must
therefore be understood in this variegated framework.

The formulation of a Zeno dynamics in terms of an adiabatic
theorem is powerful. Indeed one can use all the machinery of
adiabatic theorems in order to get results in this context. An
interesting extension would be to consider time-dependent
measurements
\beq
H_{\mathrm{meas}}=H_{\mathrm{meas}}(t),
\eeq
whose spectral projections $P_n=P_n(t)$ have a nontrivial time
evolution. In this case, instead of confining the quantum state to
a fixed sector, one can transport it along a given path (subspace)
$\cH_{P_n(t)}$, according to Eqs.\
(\ref{eq:mapS})-(\ref{eq:pntpn01}). One then obtains a dynamical
generalization of the process pioneered by Von Neumann in terms of
projection operators \cite{von,AV}.

\section{Example: three-level system}

\label{sec-applications}
\andy{applications}

In the present and in the following sections we will elaborate on
some examples considered in \cite{zenoreview,Napoli,zenowaseda}.
Our attention will be focused on possible applications in quantum
computation.

Reconsider (and rewrite) Peres' Hamiltonian (\ref{eq:ham3l0})
\andy{ham3l}
\beq
H_{\mathrm{3lev}} = \pmatrix{0 & \Omega & 0\cr \Omega & 0 & K
\cr 0 & K & 0} = H+ K H_{\mathrm{meas}}, \label{eq:ham3l}
\eeq
where
\beq
H=\Omega (\ket{1}\bra{2}+ \ket{2}\bra{1})=\Omega
\pmatrix{
  0 & 1 & 0 \cr
  1 & 0 & 0 \cr
  0 & 0 & 0
},
\eeq
\beq
H_{\mathrm{ meas}}= \ket{2}\bra{3}+\ket{3}\bra{2} =
\pmatrix{
  0 & 0 & 0 \cr
  0 & 0 & 1 \cr
  0 & 1 & 0
}. \label{eq:Hmeas2}
\eeq
Let us reinterpret the results of Sec.\ \ref{sec-QZEcont1} in the
light of the theorem proved in Sec.\ \ref{sec-dynamicalQZE}. As
$K$ is increased, the Hilbert space is split into three invariant
subspaces (eigenspaces of $H_{\mathrm{meas}}$)
$\cH=\bigoplus\cH_{P_n}$
\beq
\cH_{P_0}=\{ \ket{1}\}, \quad \cH_{P_1}=\{
(\ket{2}+\ket{3})/\sqrt{2}\}, \quad \cH_{P_{-1}}=\{
(\ket{2}-\ket{3})/\sqrt{2}\}, \label{eq:3sub}
\eeq
corresponding to the projections
\beq
P_0=\pmatrix{ 1 & 0 & 0 \cr 0 & 0 & 0 \cr 0 & 0 & 0 }, \quad
P_1=\frac{1}{2}\pmatrix{ 0 & 0 & 0 \cr 0 & 1 & 1 \cr 0 & 1 & 1 },
\quad P_{-1}=\frac{1}{2}\left(\begin{array}{rrr}
  0 & 0 & 0 \\
  0 & 1 & -1 \\
  0 & -1 & 1
\end{array}\right),
\eeq
with eigenvalues $\eta_0=0$ and $\eta_{\pm1}=\pm1$. The diagonal
part of the system Hamiltonian $H$ vanishes,
$H_{\mathrm{diag}}=\sum P_n H P_n=0$, and the Zeno evolution is
governed by
\beq
\label{eq:HZ3lev}
H^{\mathrm{Z}}_{\mathrm{3lev}}=H_{\mathrm{diag}}+K
H_{\mathrm{meas}}=K H_{\mathrm{meas}}=\pmatrix{0 & 0 & 0\cr 0 & 0
& K \cr 0 & K & 0} .
\eeq
Any transition between $\ket{1}$ and $\ket{2}$ is inhibited: a
watched pot never boils. This simple model has a lot of nice
features and will enable us to focus on several interesting
issues. We will therefore look in detail at its properties and
generalize them in the following sections.

\section{Zeno dynamics in a tensor-product space}

\label{sec-Zenotensor}
 \andy{Zenotensor}

In the preceding example the initial state of the apparatus
(namely the initial population of level $\ket{3}$) has a strong
influence on the free evolution of the system (levels $\ket{1}$
and $\ket{2}$). Such an influence entails also unwanted spurious
effects: the apparatus is, in some sense, ``entangled" with the
system, even if $K=0$. In other words, the evolution of the system
has an unpleasant dependence on the state of the apparatus: the
system can make Rabi transitions (between states $\ket{1}$ and
$\ket{2}$) only if the ``detector" is not excited (i.e.\ state
$\ket{3}$ is not populated). If, on the other hand, state
$\ket{3}$ is initially considerably populated, the dynamics of the
system is almost completely frozen. This is not a pleasant feature
(although one should not be too demanding for such a simple toy
model).

In a certain sense the QZE is counterintuitive in this case just
because, if the initial state is $\simeq \ket{1}$, although the
interaction strongly tends to drive the system into state
$\ket{3}$, the system remains in state $\ket{1}$. On the other
hand, one wonders whether such an effect would take place if the
initial state of the apparatus would have little or no influence
on the system evolution. This would give a better picture of the
QZE: the interaction Hamiltonian should be chosen in such a way
that the measured system modifies the state of the apparatus
without significant back reaction. In other words, the dynamics of
the system should not depend on the state of the apparatus: the
apparatus should simply ``register" the system evolution
(performing a spectral decomposition \cite{Wigner63,PN94}) without
``affecting" it.

The most convenient scheme for describing such a better notion of
measurement is to consider the system and the detector as two
different degrees of freedom living in \textit{different} Hilbert
spaces $\cH_{\mathrm{s}}$ and $\cH_{\mathrm{d}}$, respectively.
The combined total system evolves therefore in the tensor-product
space
\beq
\cH=\cH_{\mathrm{s}}\otimes\cH_{\mathrm{d}}
\label{eq:tensorprod}
\eeq
according to the generic Hamiltonian
\beq
H_{\mathrm{prod}}= H_{\mathrm{s}} \otimes 1_{\mathrm{d}} +
1_{\mathrm{s}} \otimes H_{\mathrm{d}} + K H_{\mathrm{meas}} .
\label{eq:Hprod}
\eeq

The theorem of Sec.\ \ref{sec-dyntheo} is naturally formulated in
the total Hilbert space $\cH$, without taking into account its
possible tensor-product decomposition. On the other hand, one
would like to shed more light on the Zeno evolution of the system
and the apparatus in their respective spaces, $\cH_{\mathrm{s}}$
and $\cH_{\mathrm{d}}$, in order to understand whether there is
such a simple prescription as (\ref{eq:Zenogenerator}) and
(\ref{eq:measinterbis}) in each component space.

\subsection{Three-level system revisited}
 \label{sec-Peresqubits}
 \andy{Peresqubits}

Let us first reconsider the example of Sec.\
\ref{sec-applications}. The (3-dimensional) Hamiltonian
(\ref{eq:ham3l}) is expressed in terms of a direct-sum Hilbert
space $\cH=\cH_{\mathrm{s}}\oplus\cH_{\mathrm{d}}$, but can be
readily reformulated in terms of the tensor-product Hilbert space
of two 2-dimensional Hilbert spaces, i.e.\ in terms of two coupled
qubits $\kket{i}_{\mathrm{s}}$ and $\kket{i}_{\mathrm{d}}$
$(i=0,1)$, as
\beq
\label{eq:ham3lprod}
H_{\mathrm{3lev}}=\Omega\; \sigma_{1\mathrm{s}} \otimes
P_{0\mathrm{d}} + K\; P_{1\mathrm{s}}\otimes \sigma_{1\mathrm{d}},
\eeq
where $\sigma_1=\kket{0}\bbra{1}+\kket{1}\bbra{0}$ and
$P_i=\kket{i}\bbra{i}$. Indeed, it is easy to show that, by
identifying
\beq
\ket{1}=\kket{00}, \quad \ket{2}=\kket{10}, \quad \ket{3}=\kket{11},
\eeq
where
$\kket{ij}=\kket{i}_{\mathrm{s}}\otimes\kket{j}_{\mathrm{d}}$, the
Hamiltonian (\ref{eq:ham3lprod}) becomes the Hamiltonian
(\ref{eq:ham3l}). The fourth available state $\ket{4}=\kket{01}$
of the tensor-product space is idle and decouples from the others.

The unwanted features of the apparatus discussed at the beginning
of this section are apparent in Eq.\ (\ref{eq:ham3lprod}): the
system-Hamiltonian $\Omega \sigma_{1\mathrm{s}}$ is effective
\textit{only} if the detector is in state $\kket{0}_{\mathrm{d}}$.
It is also apparent that the minimal modification that fits the
general form (\ref{eq:Hprod}) is simply
\beq
\label{eq:ham3lprod1} H_{\mathrm{3lev}}'=\Omega\;
\sigma_{1\mathrm{s}} \otimes 1_{\mathrm{d}} + K\;
P_{1\mathrm{s}}\otimes \sigma_{1\mathrm{d}}.
\eeq
Note that $H_{\mathrm {meas}}=P_{1\mathrm{s}}\otimes
\sigma_{1\mathrm{d}}= \ket{2}\bra{3}+\ket{3}\bra{2}$ is not
changed, whence its three eigenspaces are still
\barr
\cH_{P_0}&=&\{ \ket{1}, \ket{4}\}=\{ \kket{10}, \kket{11}\},
\nonumber\\
\cH_{P_1}&=&\{ (\ket{2}+\ket{3})/\sqrt{2}\}=
\{\kket{1}_{\mathrm{s}}\otimes \kket{+x}_{\mathrm{d}}\},
\nonumber\\
\cH_{P_{-1}}&=&\{(\ket{2}-\ket{3})/\sqrt{2}\}=\{\kket{1}_{\mathrm{s}}\otimes
\kket{-x}_{\mathrm{d}}\}
\label{eq:3sub12}
\earr
[remember that the enlarged product space contains also a fourth
idle state $\ket{4}=\kket{01}$], with eigenprojections
\beq
P_0=P_{0\mathrm{s}}\otimes 1_{\mathrm{d}}, \qquad
P_1=P_{1\mathrm{s}}\otimes P_{+x\mathrm{d}}, \qquad
P_{-1}=P_{1\mathrm{s}}\otimes P_{-x\mathrm{d}},
\eeq
where $\kket{\pm x}=[\kket{0}\pm\kket{1}]/\sqrt{2}$ and $P_{\pm
x}=\kket{\pm x}\bbra{\pm x}$. As a consequence, the Zeno evolution
is the same as before
\beq
H_{\mathrm{3lev}}^{\prime\;\mathrm{Z}}= \sum_{n=-1}^{+1} P_n
H_{\mathrm{3lev}} P_n=K\; P_{1\mathrm{s}}\otimes
\sigma_{1\mathrm{d}}= K
H_{\mathrm{meas}}=H_{\mathrm{3lev}}^{\mathrm{Z}},
\eeq
see (\ref{eq:HZ3lev}). This proves that the answer to the implicit
question at the beginning of this section is affirmative: it is
indeed possible to design the apparatus in such a way that its
initial state has little or no influence on the system evolution
(so that the apparatus can be properly regarded as a sort of
``pointer"); nevertheless, the measurement is as effective as
before and yields QZE.

\subsection{Two coupled qubits}
 \label{sec-generqubits}
 \andy{generqubits}

In order to understand better the role of $H_{\mathrm{meas}}$ in a
product space, we study two coupled qubits (system and detector),
living in the product space
\beq
\cH=\mathbb{C}^2\otimes \mathbb{C}^2,
\eeq
whose evolution is engendered by the Hamiltonian (\ref{eq:Hprod}),
with an interaction of the same type as (\ref{eq:ham3lprod1})
\beq
\label{eq:Hmeasprod} H_{\mathrm{meas}}=P_{1 \mathrm{s}}\otimes
V_{\mathrm{d}}.
\eeq
This describes an ideal detector, with no ``false" events: the
detector never clicks when the system is in its initial
``undecayed" state $\kket{0}_{\mathrm{s}}$.

The spectral resolution of the interaction reads
\beq
V_{\mathrm{d}} P_{\eta_n \mathrm{d}}=\eta_n P_{\eta_n \mathrm{d}},
\qquad (n=1,2) \ ,
\eeq
that is,
\beq
H_{\mathrm{meas}}=P_{1 \mathrm{s}}\otimes (\eta_1 P_{\eta_1
\mathrm{d}}+\eta_2 P_{\eta_2 \mathrm{d}}),
\eeq
where the two eigenvalues $\eta_1$ and $\eta_2$ are not
necessarily different and nonvanishing. Therefore, the Hilbert
space is at most split into three Zeno subspaces: a
two-dimensional one, corresponding to $\eta_0=0$,
\beq
H_{\mathrm{meas}} P_0=0, \qquad P_0=P_{0 \mathrm{s}} \otimes
1_{\mathrm{d}},
\eeq
and two one-dimensional ones
\beq
\label{eq:HmPn} H_{\mathrm{meas}} P_{n} = \eta_n P_n, \qquad
P_n=P_{1 \mathrm{s}} \otimes P_{\eta_n \mathrm{d}},\quad (n=1,2)
\eeq
corresponding to $\eta_1$ and $\eta_2$. There are three different
cases.

\subsubsection{Nondegenerate case
$0=\eta_0\neq\eta_1\neq\eta_2\neq\eta_0$}

In the nondegenerate case $0=\eta_0\neq\eta_1\neq\eta_2\neq\eta_0$
the apparatus is able to distinguish the three subspaces and the
total Hilbert space is split into
\barr
& &\cH=\cH_0\oplus\cH_1\oplus\cH_2
\nonumber\\
& &\cH_0=\{\kket{00},\kket{01}\}, \quad
\cH_1=\{\kket{1}_s\otimes\kket{\eta_1}_d\}, \quad
\cH_2=\{\kket{1}_s\otimes\kket{\eta_2}_d\}. \;\;\;\;
\earr
Therefore (\ref{eq:Hprod}) yields (for large $K$) the Zeno
Hamiltonian
\barr
H^{\mathrm{Z}}_{\mathrm{prod}}&=&\sum_{n=0}^2 P_n
H_{\mathrm{prod}} P_n
\nonumber\\
& =& (P_{0\mathrm{s}} H_{\mathrm{s}} P_{0\mathrm{s}}+
P_{1\mathrm{s}}H_{\mathrm{s}} P_{1\mathrm{s}})\otimes
1_{\mathrm{d}}
\nonumber\\
& &+ P_{0\mathrm{s}}\otimes H_{\mathrm{d}} +
P_{1\mathrm{s}}\otimes (P_{\eta_1 \mathrm{d}} H_{\mathrm{d}}
P_{\eta_1 \mathrm{d}}+P_{\eta_2 \mathrm{d}} H_{\mathrm{d}}
P_{\eta_2 \mathrm{d}})+ K H_{\mathrm{meas}}. \;\;\;\;\;
\label{eq:HprodZeno}
\earr
One should notice that the resulting effect on the system
Hamiltonian $H_{\mathrm{s}}\otimes 1_{\mathrm{d}}$ is simply the
replacement
\beq
H_{\mathrm{s}}\rightarrow
H_{\mathrm{s}}^{\mathrm{Z}}=P_{0\mathrm{s}} H_{\mathrm{s}}
P_{0\mathrm{s}}+ P_{1\mathrm{s}}H_{\mathrm{s}} P_{1\mathrm{s}},
\label{eq:HsZ}
\eeq
satisfying our expectations (QZE). On the other hand, for the
detector Hamiltonian $1_{\mathrm{s}} \otimes H_{\mathrm{d}}$ such
a simple replacement is not possible, for the resulting dynamics
is entangled. This is a consequence of the fact that the
interaction is able to distinguish between different detector
states [$P_n$ in (\ref{eq:HmPn})] in the subspace of the decay
products $P_{1\mathrm{s}}\otimes 1_{\mathrm{d}}$. If the
interaction Hamiltonian (\ref{eq:Hmeasprod}) commutes with the
detector Hamiltonian,
\beq
[V_{\mathrm{d}}, H_{\mathrm{d}}]=0,
\eeq
then the above-mentioned entanglement does not occur, for the
detector Hamiltonian $1_s\otimes H_{\mathrm{d}}$ remains
unchanged. In such a case, if $H_{\mathrm{d}}$ is nondegenerate,
i.e.\ if it is not proportional to the identity operator
$1_{\mathrm{d}}$, then $V_{\mathrm{d}}$ is not a good measurement
Hamiltonian. Indeed, for any value of the coupling constant $K$,
the detector qubit does not move and remains in its initial
pointer eigenstate (eigenstate of $H_{\mathrm{d}}$). Nevertheless,
the QZE is still effective. See also the next case.

On the other hand, a good detector has an interaction Hamiltonian
$V_{\mathrm{d}}$ which is a complementary observable
\cite{von,Schwinger59} of its free Hamiltonian $H_{\mathrm{d}}$.
For example, if we set, without loss of generality,
$H_{\mathrm{d}}= b \sigma_{3\mathrm{d}}$, the interaction should
be $V_{\mathrm{d}}= \sigma_{1\mathrm{d}}$ (or
$V_{\mathrm{d}}=\sigma_{2\mathrm{d}}$). In such a case, the
diagonal part of an observable with respect to the other vanishes,
i.e.\ $P_{\eta_1 \mathrm{d}} H_{\mathrm{d}} P_{\eta_1
\mathrm{d}}+P_{\eta_2 \mathrm{d}} H_{\mathrm{d}} P_{\eta_2
\mathrm{d}}=0$, and the Zeno Hamiltonian (\ref{eq:HprodZeno})
reads
\barr
H^{\mathrm{Z}}_{\mathrm{prod}}= (P_{0\mathrm{s}} H_{\mathrm{s}}
P_{0\mathrm{s}}+ P_{1\mathrm{s}}H_{\mathrm{s}}
P_{1\mathrm{s}})\otimes 1_{\mathrm{d}} + P_{0\mathrm{s}}\otimes
H_{\mathrm{d}} + K H_{\mathrm{meas}}.
\label{eq:HprodZeno1}
\earr
It is therefore apparent that, in the case of a good detector, not
only the system evolution, but also the detector evolution is
hindered (QZE). Indeed, in the large-$K$ limit, if the system
qubit starts (and remains) in $\kket{0}_{\mathrm{s}}$, then the
pointer qubit is frozen as well in one of its eigenstates (the
eigenstates of $H_{\mathrm{d}}$).

\subsubsection{Degenerate interaction $0=\eta_0\neq\eta_1=\eta_2$}

In this case there are only two projections
\beq
P_0=P_{0 \mathrm{s}} \otimes 1_{\mathrm{d}}, \quad \tilde
P_1=P_1+P_2=P_{1\mathrm{s}}
\otimes 1_{\mathrm{d}}
\eeq
and two 2-dimensional Zeno subspaces
\barr
& &\cH=\cH_0\oplus\tilde\cH_1
\nonumber\\
& &\cH_0=\{\kket{00},\kket{01}\}, \quad
\tilde\cH_1=\{\kket{10}+\kket{11}\}.
\earr
The Zeno Hamiltonian reads
\barr
H^{\mathrm{Z}}_{\mathrm{prod}}&=& P_0 H_{\mathrm{prod}} P_0 +
\tilde P_1 H_{\mathrm{prod}} \tilde P_1
\nonumber\\
& =& (P_{0\mathrm{s}} H_{\mathrm{s}} P_{0\mathrm{s}}+
P_{1\mathrm{s}} H_{\mathrm{s}} P_{1\mathrm{s}})\otimes
1_{\mathrm{d}} + 1_{\mathrm{s}}\otimes H_{\mathrm{d}} + K
H_{\mathrm{meas}}
\earr
and the QZE occurs again according to (\ref{eq:HsZ}), leaving the
detector Hamiltonian unaltered and without creating entanglement.
Notice that in this case the interaction (\ref{eq:Hmeasprod})
reduces to
\beq
H_{\mathrm{meas}}=\eta_1 P_{1 \mathrm{s}}\otimes 1_{\mathrm{d}}
\eeq
and does not yield an evolution of the detector qubit. In spite of
this, the Hilbert space is split into two Zeno subspaces and a QZE
takes place. This happens because some information is stored in
the phase of the detector qubit.

\subsubsection{Imperfect measurement
$0=\eta_0=\eta_1\neq\eta_2$}

In this last situation, there are again two projections,
\beq
\tilde P_0=P_0+P_1=P_{0 \mathrm{s}} \otimes 1_{\mathrm{d}} +P_{1
\mathrm{s}} \otimes P_{\eta_1 \mathrm{d}} , \quad
P_2=P_{1\mathrm{s}} \otimes P_{\eta_2 \mathrm{d}} ,
\eeq
and two Zeno subspaces,
\barr
& &\cH=\tilde \cH_0\oplus\cH_2
\nonumber\\
& &\tilde\cH_0=\{\kket{00},\kket{01},
\kket{1}_{\mathrm{s}}\otimes\kket{\eta_1}_{\mathrm{d}}\}, \quad
\cH_2=\{\kket{1}_{\mathrm{s}}\otimes\kket{\eta_2}_{\mathrm{d}} \}:
\earr
a 3-dimensional one, corresponding to the eigenvalue $\eta_0=0$
and a 1-dimensional one, corresponding to $\eta_2\neq 0$. However,
in this case the measuring interaction is not able to perform a
clear-cut distinction between the initial state
$\kket{0}_{\mathrm{s}}$ of the system and its decay product
$\kket{1}_{\mathrm{s}}$, i.e.\ it yields an \textit{imperfect}
measurement.

The Zeno Hamiltonian reads
\barr
H^{\mathrm{Z}}_{\mathrm{prod}}&=& \tilde P_0 H_{\mathrm{prod}}
\tilde P_0 + P_2 H_{\mathrm{prod}} P_2
\nonumber\\
& =& H_{\mathrm{s}}\otimes P_{\eta_1 \mathrm{d}} +
(P_{0\mathrm{s}} H_{\mathrm{s}} P_{0\mathrm{s}}+
P_{1\mathrm{s}}H_{\mathrm{s}} P_{1\mathrm{s}})\otimes P_{\eta_2
\mathrm{d}}
\nonumber\\
& & + P_{0 \mathrm{s}}\otimes H_{\mathrm{d}} +
P_{1\mathrm{s}}\otimes (P_{\eta_1 \mathrm{d}} H_{\mathrm{d}}
P_{\eta_1 \mathrm{d}}+P_{\eta_2 \mathrm{d}} H_{\mathrm{d}}
P_{\eta_2 \mathrm{d}}) + K H_{\mathrm{meas}}. \;\;\;\;
\earr
Notice that $H^{\mathrm{Z}}_{\mathrm{prod}}$ displays an
interesting symmetry between the system and the apparatus. The
origin of this symmetry is apparent by looking at the interaction
Hamiltonian $H_{\mathrm{meas}}$:
\beq
H_{\mathrm{meas}}=\eta_1 P_{1 \mathrm{s}}\otimes P_{\eta_2
\mathrm{d}} .
\eeq
A partial QZE is still present. In fact, the evolution of the
system is frozen only if the detector is in state
$\kket{\eta_2}_{\mathrm{d}}$, while it is not hindered if the
latter is in state $\kket{\eta_1}_{\mathrm{d}}$ (and a similar
situation holds for the detector evolution).

The three cases analyzed in this subsection are paradigms for
examining the rich behavior of the Zeno dynamics engendered by
Hamiltonian (\ref{eq:Hprod}) in a generic tensor-product space
(\ref{eq:tensorprod}). In particular, one can show that, by
considering a good detector (whose free and interaction
Hamiltonians, $H_{\mathrm{d}}$ and $V_{\mathrm{d}}$, are two
generic complementary observables \cite{Schwinger01}), the Zeno
Hamiltonian (\ref{eq:HprodZeno1}) admits a straightforward natural
generalization to the $N$-dimensional case. We shall elaborate
further on this issue in a future paper.

\section{A watched cook can freely watch a boiling pot}
 \label{sec-cook}
 \andy{cook}

Let us look at another interesting model. Consider
\andy{ham4l}
\beq
H_{\mathrm{4lev}} =\Omega \sigma_1+ K\tau_1+K' \tau'_1 =
\pmatrix{0 & \Omega & 0 & 0 \cr \Omega & 0 & K & 0 \cr 0 & K & 0 &
K' \cr 0 & 0 & K' & 0 }, \label{eq:ham4l}
\eeq
where states $\ket{1}$ and $\ket{2}$ make Rabi oscillations,
\beq
\Omega \sigma_1=\Omega (\ket{2}\bra{1}+\ket{1}\bra{2})= \Omega
\pmatrix{0 & 1 & 0 & 0 \cr 1 & 0 & 0 & 0 \cr 0 & 0 & 0 & 0 \cr 0 &
0 & 0 & 0 } ,
\eeq
while state $\ket{3}$ ``observes" them,
\beq
K \tau_1=K (\ket{3}\bra{2}+\ket{2}\bra{3})= K \pmatrix{0 & 0& 0 &
0 \cr 0 & 0 & 1 & 0 \cr 0 & 1 & 0 & 0 \cr 0 & 0 & 0 & 0 }
\eeq
and state $\ket{4}$ ``observes" whether level $\ket{3}$ is
populated,
\beq
K' \tau'_1=K' (\ket{4}\bra{3}+\ket{3}\bra{4})= K' \pmatrix{0 & 0&
0 & 0 \cr 0 & 0 & 0 & 0 \cr 0 & 0 & 0 & 1 \cr 0 & 0 & 1 & 0 } .
\eeq
If $K \gg \Omega$ \textit{and} $K'$, then (\ref{eq:ham4l}) must be
read
\andy{ham4lnew}
\barr
H_{\mathrm{4lev}} = H + K H_{\mathrm{meas}}, \qquad
\mathrm{with}\quad
 H= \Omega
\sigma_1+ K' \tau'_1, \quad H_{\mathrm{meas}}= \tau_1,
 \label{eq:ham4lnew}
\earr
and the total Hilbert space splits into the three eigenspaces of
$H_{\mathrm{meas}}$ [compare with (\ref{eq:3sub}) and
(\ref{eq:3sub12})]:
\beq
\cH_{P_0}=\{ \ket{1}, \ket{4}\}, \quad \cH_{P_1}=\{
(\ket{2}+\ket{3})/\sqrt{2}\}, \quad \cH_{P_{-1}}=\{
(\ket{2}-\ket{3})/\sqrt{2}\}. \label{eq:3sub1}
\eeq
Moreover, $H_{\mathrm{diag}}= \sum_n P_nHP_n=0$ and the Zeno
evolution is governed by
\beq
H_{\mathrm{4lev}}^{\mathrm{Z}}= K \tau_1=\pmatrix{0 & 0 & 0 & 0
\cr 0 & 0 & K & 0 \cr 0 & K & 0 & 0 \cr 0 & 0 & 0 & 0 }.
\eeq
The Rabi oscillations between states $\ket{1}$ and $\ket{2}$ are
hindered.

On the other hand, if $K' \gg K$ and $\Omega$ (and even if $K \gg
\Omega$), then (\ref{eq:ham4l}) must be read
\andy{ham4lnnew}
\barr
H_{\mathrm{4lev}} = H + K' H_{\mathrm{meas}}, \qquad \mathrm{with}
\qquad H= \Omega \sigma_1+ K \tau_1, \quad H_{\mathrm{meas}}=
\tau'_1,
 \label{eq:ham4lnnew}
\earr
the total Hilbert space splits into the three eigenspaces of
$H_{\mathrm{meas}}$ [notice the differences with
(\ref{eq:3sub1})]:
\beq
\cH_{P'_0}=\{ \ket{1}, \ket{2}\}, \quad \cH_{P'_1}=\{
(\ket{3}+\ket{4})/\sqrt{2}\}, \quad \cH_{P'_{-1}}=\{
(\ket{3}-\ket{4})/\sqrt{2}\} \label{eq:3sub11}
\eeq
and the Zeno Hamiltonian reads
\beq
H_{\mathrm{4lev}}^{\mathrm{Z}\;\prime}=
\Omega\sigma_1+K'\tau'_1=\pmatrix{0 & \Omega & 0 & 0 \cr \Omega &
0 & 0 & 0 \cr 0 & 0 & 0 & K' \cr 0 & 0 & K' & 0 } .
\eeq
The Rabi oscillations between states $\ket{1}$ and $\ket{2}$ are
fully \textit{restored} (even if and in spite of $K \gg \Omega$)
\cite{Militello01}. A watched cook can freely watch a boiling pot.

\section{Quantum computation and decoherence-free subspaces}
 \label{sec-decfree}
 \andy{decfree}

\begin{figure}[t]
\begin{center}
\includegraphics[width=11cm]{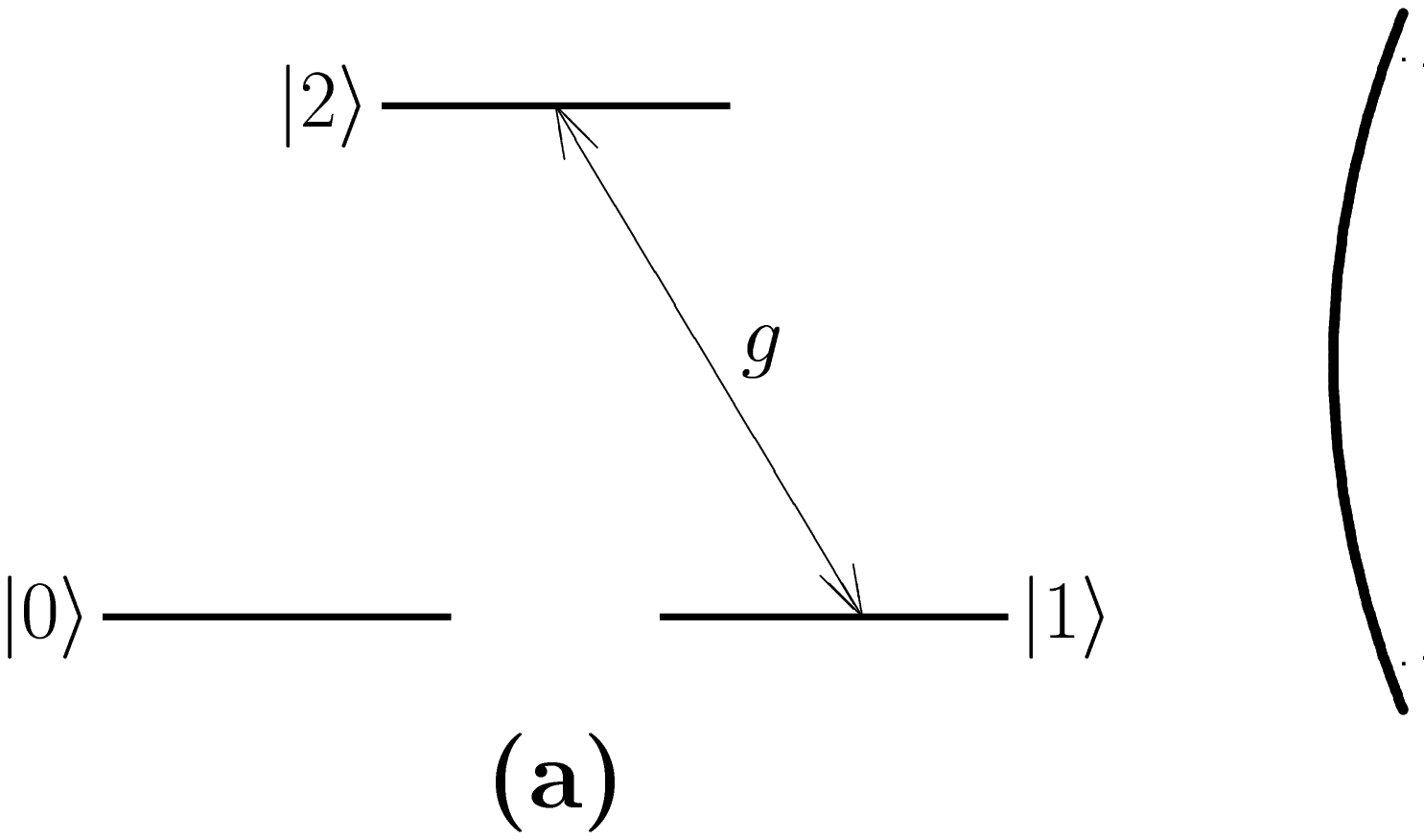}
\end{center}
\caption{Schematic view of the system described by the Hamiltonian
(\protect\ref{eq:cavity}). }
\label{fig:beige}
\end{figure}

We now look at a more realistic example, analyzing the possibility
of devising decoherence-free subspaces \cite{Viola99}, that are
relevant for quantum computation. The Hamiltonian \cite{Beige00}
\beq
\label{eq:cavity}
H_{\mathrm{meas}}=i g \sum_{i=1}^2 \left( b\;
\ket{2}_{ii}\bra{1} - b^\dagger\; \ket{1}_{ii}\bra{2}\right) - i
\kappa b^\dagger b
\eeq
describes a system of two ($i=1,2$) three-level atoms in a cavity.
The atoms are in a $\Lambda$ configuration with split ground
states $\ket{0}_i$ and $\ket{1}_i$ and excited state $\ket{2}_i$,
as shown in Fig.\ \ref{fig:beige}(a), while the cavity has a
single resonator mode $b$ in resonance with the atomic transition
1-2. See Fig.\ \ref{fig:beige}(b). Spontaneous emission inside the
cavity is neglected, but photons leak out through the nonideal
mirrors with a rate $\kappa$.

The excitation number
\beq
\cN=\sum_{i=1,2}\ket{2}_{ii}\bra{2}+b^\dagger b ,
\eeq
commutes with the Hamiltonian,
\beq
[H_{\mathrm{meas}},\cN]=0.
\eeq
Therefore we can solve the eigenvalue equation inside each
eigenspace of $\cN$ (Tamm-Duncoff sectors).

A comment is now in order. Strictly speaking, the Hamiltonian
(\ref{eq:cavity}) is non-Hermitian and we cannot directly apply
the theorem of Sec.\ \ref{sec-dyntheo}. (Notice that the proof of
the theorem heavily hinges upon the hermiticity of the
Hamiltonians and the unitarity of the evolutions.) However, we can
apply the technique outlined at the end of Sec.\ \ref{sec-QZEnH}
and enlarge our Hilbert space $\cH$, by including the photon modes
outside the cavity $a_\omega$ and their coupling with the cavity
mode $b$. The enlarged dynamics is then generated by the
\textit{Hermitian} Hamiltonian
\barr
\tilde H_{\mathrm{meas}} &=& i g \sum_{i=1}^2 \left( b\;
\ket{2}_{ii}\bra{1} - b^\dagger\; \ket{1}_{ii}\bra{2}\right)
\nonumber\\
& &+
\int d\omega\; \omega a^\dagger_\omega a_\omega +
\sqrt{\frac{\kappa}{\pi}}\int d\omega\left[a^\dagger_\omega b +
a_\omega b^\dagger\right]
\earr
and it is easy to show that the evolution engendered by $\tilde
H_{\mathrm{meas}}$, when projected back to $\cH$, is given by the
effective non-Hermitian Hamiltonian (\ref{eq:cavity}), provided
the field outside the cavity is initially in the vacuum state.
Notice that any complex eigenvalue of $H_{\mathrm{meas}}$
engenders a dissipation (decay) of $\cH$ into the enlarged Hilbert
space embedding it. On the other hand, any real eigenvalue of
$H_{\mathrm{meas}}$ generates a unitary dynamics which preserves
the probability within $\cH$. Hence it is also an eigenvalue of
$\tilde H_{\mathrm{meas}}$ and its eigenvectors are the
eigenvectors of the restriction $\tilde H_{\mathrm{meas}}|_{\cH}$.
Therefore, as a general rule, the theorem of Sec.\
\ref{sec-dyntheo} can be applied also to non-Hermitian measurement
Hamiltonians $\cH_{\mathrm{meas}}$, provided one restricts one's
attention \textit{only} to their \textit{real} eigenvalues.

The eigenspace $\cS_0$ corresponding to $\cN=0$ is spanned by four
vectors
\beq
\cS_0=\{\ket{000},\ket{001},\ket{010},\ket{011}\}, \label{eq:cN0}
\eeq
where $\ket{0j_1 j_2}$ denotes a state with no photons in the
cavity and the atoms in state $\ket{j_1}_1\ket{j_2}_2$. The
restriction of $H_{\mathrm{meas}}$ to $\cS_0$ is the null operator
\beq
H_{\mathrm{meas}}|_{\cS_0} =0,
\eeq
hence $\cS_0$ is a subspace of the eigenspace $\cH_{P_0}$ of
$H_{\mathrm{meas}}$ belonging to the eigenvalue $\eta_0=0$
\beq
\cS_0 \subset \cH_{P_0}, \qquad
H_{\mathrm{meas}} P_0= 0 .
\eeq
The eigenspace $\cS_1$ corresponding to $\cN=1$ is spanned by
eight vectors
\beq
\cS_1=\{\ket{020},\ket{002},\ket{100},\ket{110},\ket{101},\ket{021},\ket{012},\ket{111}\},
\label{eq:cN1}
\eeq
and the restriction of $H_{\mathrm{meas}}$ to $\cS_1$ is
represented by the 8-dimensional matrix
\beq
H_{\mathrm{meas}}|_{\cS_1} =\left(
\begin{array}{cccccccc}
  0 & 0 & 0 & ig & 0 & 0 & 0 & 0 \\
  0 & 0 & 0 & 0 & ig & 0 & 0 & 0 \\
  0 & 0 & -i\kappa & 0 & 0 & 0 & 0 & 0 \\
  -ig & 0 & 0 & -i\kappa & 0 & 0 & 0 & 0 \\
  0 & -ig & 0 & 0 & -i\kappa & 0 & 0 & 0 \\
  0 & 0 & 0 & 0 & 0 & 0 & 0 & ig \\
  0 & 0 & 0 & 0 & 0 & 0 & 0 & ig \\
  0 & 0 & 0 & 0 & 0 & -ig & -ig & -i\kappa
\end{array}\right) .
\eeq
It is easy to prove that the eigenvector
$(\ket{021}-\ket{012})/\sqrt{2}$ has eigenvalue $\eta_0=0$ and all
the other eigenvectors have eigenvalues with negative imaginary
parts. Moreover, all restrictions $H_{\mathrm{meas}}|_{\cS_n}$
with $n>1$ have eigenvalues with negative imaginary parts. Indeed
they are spanned by states containing at least one photon, which
dissipates through the nonideal mirrors, according to $-i\kappa
b^\dagger b$ in (\ref{eq:cavity}). The only exception is state
$\ket{0,2,2}$ of $\cS_2$, but also in this case it easy to prove
that all eigenstates of $H_{\mathrm{meas}}|_{\cS_2}$ dissipate. In
conclusion, blending these results with (\ref{eq:cN0}), one infers
that the eigenspace $\cH_{P_0}$ of $H_{\mathrm{meas}}$ belonging
to the eigenvalue $\eta_0=0$ is 5-dimensional and is spanned by
\andy{decfree}
\beq
\cH_{P_0}=\{\ket{000},\ket{001},\ket{010},\ket{011},(\ket{021}-\ket{012})/\sqrt{2}\},
\label{eq:decfree}
\eeq
If the coupling $g$ and the cavity loss $\kappa$ are sufficiently
strong, any other weak Hamiltonian $H$ added to (\ref{eq:cavity})
reduces to $P_0 H P_0$ and changes the state of the system only
\textit{within} the decoherence-free subspace (\ref{eq:decfree}).
This corroborates the conclusions of \cite{Beige00} and completely
characterizes the decoherence-free subspaces in this example. This
could be relevant for practical applications.

\section{Spontaneous decay in vacuum}
 \label{sec-vacdec}
 \andy{vacdec}

Our last example deals with spontaneous decay in vacuum. Let
\andy{hamqc}
\beq
H_{\mathrm{decay}}= H+ K H_{\mathrm{meas}}= \pmatrix{0 &
\tau_{\mathrm{Z}}^{-1} & 0\cr \tau_{\mathrm{Z}}^{-1} & -i 2/
\tau_{\mathrm{Z}}^2 \gamma & K \cr 0 & K & 0}. \label{eq:hamqc}
\eeq
This describes the spontaneous emission $\ket{1}\to\ket{2}$ of a
system into a (structured) continuum, while level $\ket{2}$ is
resonantly coupled to a third level $\ket{3}$ \cite{zenoreview}.
The quantity $\gamma$ represents the decay rate to the continuum
and $\tau_{\mathrm{Z}}$ is the Zeno time (convexity of the initial
quadratic region). This case is also relevant for quantum
computation, if one is interested in protecting a given subspace
(level $\ket{1}$) from decoherence by inhibiting spontaneous
emission. A somewhat related example is considered in
\cite{Agarwal01}. Model (\ref{eq:hamqc}) is also relevant for some
examples analyzed in \cite{Viola99} and \cite{Beige00}, but we
will not elaborate on this point here.

Notice that, in a certain sense, this situation is complementary
to that in (\ref{eq:cavity}); here the measurement Hamiltonian
$H_{\mathrm{meas}}$ is Hermitian, while the system Hamiltonian $H$
is not. Again, one has to enlarge the Hilbert space, as in Secs.\
\ref{sec-QZEnH} and \ref{sec-decfree}, apply the theorem to the
dilation and project back the Zeno evolution. As a result one can
simply apply the theorem to the original Hamiltonian
(\ref{eq:hamqc}), for in this case $H_{\mathrm{meas}}$ has a
complete set of orthogonal projections that univocally defines a
partition of $\cH$ into Zeno subspaces. We shall elaborate further
on this interesting aspect in a future paper.

As the Rabi frequency $K$ is increased, one is able to hinder
spontaneous emission from level $\ket{1}$ (to be ``protected" from
decay/decoherence) to level $\ket{2}$. However, in order to get an
effective ``protection" of level $\ket{1}$, one needs $K>
1/\tau_{\mathrm{Z}}$. More to this, if the initial state $\ket{1}$
has energy $\omega_1\neq 0$, an inverse Zeno effect takes place
\cite{zenowaseda} and the requirement for obtaining QZE becomes
even more stringent
\cite{heraclitus}, yielding $K>1/\tau_{\mathrm{Z}}^2\gamma$. Both
these conditions can be very demanding for a real system subject
to dissipation \cite{zenoreview,heraclitus,Napoli}. For instance,
typical values for spontaneous decay in vacuum are $\gamma\simeq
10^9$s$^{-1}$, $\tau_{\mathrm{Z}}^2\simeq 10^{-29}$s$^2$ and
$1/\tau_{\mathrm{Z}}^2\gamma\simeq 10^{20}$s$^{-1}$
\cite{hydrogen}.

We emphasize that the example considered in this subsection is not
to be regarded as a toy model. The numerical figures we have given
are realistic and the Hamiltonian (\ref{eq:hamqc}) is a good
approximation at short (for the physical meaning of ``short", see
\cite{zenoreview,heraclitus,Napoli}) and intermediate times.

\section{Conclusions}
\label{sec-concl}

The usual formulation of the QZE (and IZE) hinges upon the notion
of pulsed measurements, according to von Neumann's projection
postulate. However, as we pointed out, a ``measurement" is nothing
but an interaction with an external system (another quantum
object, or a field, or simply another degree of freedom of the
very system investigated), playing the role of apparatus. This
remark enables one to reformulate the Zeno effects in terms of a
(possibly strong or finely-tuned) coupling to an external agent
and to cast the quantum Zeno evolution in terms of an adiabatic
theorem. We have analyzed several examples, which might lead to
interesting applications. Among these, we have considered in some
detail the possibility of tailoring the interaction so as to
obtain decoherence-free subspaces, useful also for quantum
computation.

\end{document}